\documentclass{article} 
\usepackage[utf8]{inputenc}
\usepackage[T1]{fontenc}

\usepackage{times}
 \newcommand{\RNum}[1]{\uppercase\expandafter{\romannumeral #1\relax}}

\usepackage{amsmath,amsfonts,bm}

\def\eqref#1{equation~\ref{#1}}

\def\1{\bm{1}}

\DeclareMathAlphabet{\mathsfit}{\encodingdefault}{\sfdefault}{m}{sl}
\SetMathAlphabet{\mathsfit}{bold}{\encodingdefault}{\sfdefault}{bx}{n}

\usepackage{hyperref}
\hypersetup{hidelinks}
\usepackage{url}
\usepackage{bbm}  
\usepackage{subcaption}
\usepackage[table]{xcolor}
\usepackage{amsthm}
\usepackage{amssymb}  

\usepackage{comment}
\newcount\Comments  
\newcommand{\kibitz}[2]{\ifnum\Comments=1{\color{#1}{#2}}\fi}
\definecolor{english}{rgb}{0.0, 0.5, 0.0}

\newif\ifmarkrev
\markrevfalse
\newcommand{\rev}[1]{\ifmarkrev\begingroup\color{red}#1\endgroup\else#1\fi}

\usepackage{amsmath} 
\usepackage{graphicx}
\graphicspath{{../}{./}}  
\usepackage{hyperref}
\usepackage{natbib}
\usepackage{float}
\usepackage{geometry}
\usepackage{pgfplots}
\usepackage[framemethod=TikZ]{mdframed}
\usepackage{enumitem}
\usepackage{subcaption}
\usepackage[skins,breakable]{tcolorbox}  
\usepackage{setspace}
\usepackage{textcomp}
\usepackage{tabularx}
\usepackage{array}
\usepackage{tikz}
\usepackage{multirow}
\usetikzlibrary{arrows,positioning, automata,arrows.meta, shapes.geometric, decorations.pathreplacing}
\usepackage{booktabs} 
\title{\vspace*{-1cm} \rev{LLM Bidders Preserve the Mechanism-Level Orderings of Human Bidders} \thanks{
  \footnotesize
This research was made possible by a generous grant from Dropbox Inc. Thanks to Parker Whitfill and Kobi Gal for their helpful feedback. Author's contact information, code, and data will be available at https://github.com/KeHang-Zhu/llm-auction. Both Anand V. Shah and Kehang Zhu contributed equally to this work. John J. Horton is a co-founder of Expected Parrot Inc., which develops generative AI models for market research applications. The views expressed herein are those of the authors and do not necessarily reflect the views of the National Bureau of Economic Research.
}
}

\author{
  Anand Shah$^{*}$\\
  MIT \\
  \and
  Kehang Zhu$^{*}$\\
  Harvard  \\
  \and
  Yanchen Jiang \\
  Google Research \\
  \and
  Jeffrey G. Wang \\
  OpenAI \\
  \and
  Arif K. Dayi \\
  Harvard \\
  \and
  John J. Horton \\
  MIT \& NBER \\
  \and
  David C. Parkes \\
  Harvard \\
}
\date{
Initial draft: May 30, 2024 \\
Current version: September 20, 2026
}

\usepackage{amsmath}
\usepackage{graphicx}
\usepackage{hyperref}
\usepackage{natbib}
\usepackage{float}
\usepackage{geometry}
\usepackage{pgfplots}
\usepackage[framemethod=TikZ]{mdframed}
\usepackage{enumitem}
\usepackage{subcaption}
\usepackage[skins,breakable]{tcolorbox}  
\usepackage{setspace}
\usepackage{textcomp}
\usepackage{tabularx}
\usepackage{array}
\usepackage{tikz}
\usepackage{multirow}
\usetikzlibrary{arrows,positioning, automata,arrows.meta, shapes.geometric, decorations.pathreplacing}
\usepackage{color-edits}
\addauthor{ma}{blue}
\usepackage{soul}
\usepackage{listings}
\usepackage{colortbl} 

\begin{document}

\maketitle
\begin{abstract}
\rev{Training on vast amounts of human-generated data has motivated growing interest in using large language models (LLMs) to simulate human behavior. We ask which features of human behavior general-purpose models preserve when used out of the box in auctions, where multiple bidders interact under explicit rules and incentives. We evaluate five LLMs across seven laboratory settings against human benchmarks reconstructed from published experiments, with uncertainty bands for the private-value comparisons. Our main focus is on three large models without extended test-time reasoning: GPT-4o, Claude~3.5 Haiku, and Gemini~2.0 Flash. LLM and human deviations from theory differ in magnitude and often in direction: humans overbid in second-price auctions, whereas most models that deviate underbid. Surprisingly, without task-specific fine-tuning or calibration to human bids, the three non-reasoning large models robustly preserve key orderings of auction formats by deviation from theory. First-price auctions are harder than second-price, and ascending clocks reduce deviations relative to sealed bids wherever data are adequate. Kendall's $\tau_b$ between the human and GPT-4o difficulty rankings is $0.60$ and positive in every joint bootstrap draw. The reasoning model bids almost at equilibrium in the observed private-value settings, leaving little variation in errors to compare; the small model's large errors yield an inverted ranking. All five models nevertheless reproduce the stronger first-price winner's curse. Clock framing improves bidding for two of the three non-reasoning large models, and GPT-4o recovers the ordering of last-minute bidding across closing rules in an eBay-style marketplace.}
\end{abstract}
\newpage
\section{Introduction} \label{sec:introduction}

\rev{Training on vast amounts of human-generated data has motivated growing interest in using large language models (LLMs) to simulate human behavior \citep{aher2023using, horton2023large, manning2024automated}.
We ask which features of human behavior general-purpose models preserve when used out of the box in auctions, where multiple bidders interact under explicit rules and incentives.
A mechanism designer choosing between two auction formats needs to know how bidders will actually play them, not only what theory prescribes.
Decades of laboratory experiments show that the two can differ systematically: people bid above value in second-price auctions, shade too little in first-price auctions, fall prey to the winner's curse when values are common, and play the same strategy-proof rule far more truthfully when it is implemented as an ascending clock than as a sealed bid \citep{kagel1993independent, kagel1986winner, li2017obviously, breitmoser2022obviousness}.
Establishing each of these facts required laboratory experiments with paid human subjects, which are costly to run and slow to extend to new designs.
LLMs can be prompted to bid in the same auctions at a small fraction of the cost, which raises a practical question: can simulated bidders inform the designer's judgment, and if so, which features of human behavior do they preserve?\footnote{An early version of this paper was presented at a workshop at NeurIPS 2023.}
This paper answers that question for auctions.
We distinguish three features of behavior that a simulation could preserve: the \emph{magnitude} of the deviation from theory, its \emph{direction} (under- or over-bidding), and the \emph{ordering} of mechanisms by how far bidders deviate, together with the signs of the comparative statics between them.
The three tiers are progressively less demanding, and the third is the one a designer comparing candidate mechanisms actually uses.}

\rev{Auctions are a demanding benchmark for simulated bidders because the human evidence is unusually sharp.
Theory gives exact predictions: a Bayes--Nash bid function in first-price auctions and dominant-strategy truth-telling in second-price and clock formats \citep{myerson1981optimal, krishna2009auction}.
The experimental literature documents not only that people deviate from these predictions but by how much more they deviate in some mechanisms than in others \citep{kagel2020handbook}: bidding above the risk-neutral equilibrium in first-price auctions \citep{coppinger1980incentives, cox1988theory}, bidding above value in second-price auctions \citep{kagel1993independent}, winner's-curse losses in common-value auctions \citep{bazerman1983won, kagel1986winner, kagel1987information}, more truthful play in ascending-clock than in sealed-bid implementations of the same rule \citep{li2017obviously, breitmoser2022obviousness}, and last-minute bidding under hard-close rules in online marketplaces \citep{roth2002last, ockenfels2006late}.
This body of evidence supplies an ordering of behavioral difficulty across mechanisms, which is exactly the object a designer needs and exactly what a simulation can be tested against.
A growing literature evaluates LLMs in strategic settings against theoretical optimality; we instead evaluate them against these human regularities, with theory as the reference point rather than the criterion.}

\rev{The benchmark requires human results that are comparable across mechanisms, and the literature does not report them in that form.
Published studies report moments rather than micro data, such as overbidding shares, bid-on-value regressions, and mean absolute deviations, from experiments that differ in the number of bidders, value supports, and repetition.
We treat this as an \emph{integrated experimentation} problem.
For each source we fit a simple mixture model of bidding to the reported moments, simulate bids from the fitted model, and compute one deviation metric, the scaled mean absolute deviation from the theoretical benchmark (SMAD), that we also compute for LLM bids.
Because the fitted parameters are uncertain, we carry that uncertainty forward as reconstruction bands and use the human anchors only for orderings, directions, and comparative statics.
On the simulation side we hold the prompt fixed: every bidder receives an induced value and a single objective, to maximize expected profit, following the induced-value tradition of experimental economics \citep{smith1976experimental}.
What varies is the mechanism (first-, second-, and third-price sealed bids, open and closed ascending clocks), the information structure (independent, affiliated, and common values), and the model.
Our main focus is on three large models without extended test-time reasoning: GPT-4o, Claude~3.5 Haiku, and Gemini~2.0 Flash. We include GPT-5-mini, a reasoning model that allocates extended test-time computation to each decision, to ask whether more accurate equilibrium play also makes a model a better simulator of human behavior. The small open-weights model Llama-3-8B-Instruct provides a comparison with more limited capabilities.
We use all five models out of the box, without task-specific fine-tuning or calibration to human bids. This complements work on models specifically trained to predict human behavior, such as Centaur, which is fine-tuned on a large collection of human experimental choices \citep{binz2025foundation}.
GPT-4o additionally serves as the reference model for the deeper distributional diagnostics, because it remains a widely used and well-studied model for strategic and economic tasks \citep{horton2023large, manning2024automated}.}

\rev{The results separate cleanly by tier.
Magnitudes do not match: no model's SMAD falls within the human reconstruction band in more than two of the five private-value formats; GPT-4o deviates more than humans in five of seven formats and far less in the clock formats, GPT-5-mini plays the equilibrium almost exactly, and Llama-3-8B deviates several times more than humans everywhere.
Directions do not match: human bidders in second-price auctions predominantly bid above value \citep{kagel1993independent, li2017obviously}, whereas GPT-4o, Gemini, and Llama predominantly bid below value, GPT-5-mini bids at value, and only Claude~3.5 Haiku errs in the human direction.
Surprisingly, despite receiving no task-specific training to mimic human bidders, the non-reasoning large models preserve key orderings: first-price auctions are harder than second-price for GPT-4o, Claude, and Gemini as they are for humans, the clock implementation reduces deviations for every model with adequate clock data, and the first-price winner's curse exceeds the second-price curse for all five models.
Kendall's $\tau_b$ between the human and GPT-4o difficulty rankings over five formats is $0.60$ and positive in every joint bootstrap draw, and the ordering is robust to the sampling temperature (Appendix~\ref{app:temperature}).
The ranking is uninformative for GPT-5-mini, whose cells all sit at the benchmark, and inverted for Llama.}

\rev{These comparisons show why more accurate equilibrium play need not make a model a better simulator of human behavior.
GPT-5-mini bids almost at equilibrium in the observed private-value settings, leaving little variation in errors with which to compare human difficulty across formats; it records no usable clock bids.
Llama-3-8B makes large errors whose ordering is opposite to the human one.
The non-reasoning large models instead make systematic errors that vary across mechanisms in a similar order to human errors, even when their magnitudes and directions differ.
Thus, accuracy relative to theory and fidelity to human behavior are distinct criteria.}

\rev{We then use the ordinal result the way a designer would.
First as a prediction: describing a sealed-bid second-price auction as a clock raises truthful play among human subjects \citep{breitmoser2022obviousness}, so it should raise truthful play for the non-reasoning large models.
It does for GPT-4o and Claude~3.5 Haiku, and it fails only for Gemini~2.0 Flash, whose baseline play in this format is already close to truthful, which is where the argument predicts no gain.
Second in a market structure documented by field data rather than by laboratory experiments: in an eBay-style proxy-bidding marketplace, LLM bidders place $63\%$ of decisive bids in the last scheduled period under a hard close and $7\%$ under a soft-close extension rule, reproducing the ordering of last-minute bidding across closing rules observed on eBay and early Amazon \citep{roth2002last, ockenfels2006late}.}

\rev{The paper makes three contributions.
It provides a benchmark that converts published human auction moments into cross-mechanism comparable quantities with explicit reconstruction uncertainty, so that any simulated bidder can be scored against decades of experimental evidence.
It provides a three-tier evaluation that separates what a simulation can be read for from what it cannot, and applies it to five models across seven formats.
And it turns the result into a concrete recommendation: LLM panels can screen candidate mechanisms and presentations for cross-mechanism orderings and the signs of comparative statics, but they should not be read for the magnitude or direction of human deviations. By an \emph{LLM panel}, we mean a set of models evaluated separately on the same auction settings, so that agreement across models can be checked. Our results support using the non-reasoning large models for this purpose; the reasoning and small models show why the panel's ability to reproduce human orderings must be validated.
Our work sits within the literature that uses LLMs as simulated agents in economics and social science \citep{aher2023using, park2023generative, horton2023large, manning2024automated}, a literature that also documents the limits of LLMs as stand-ins for human respondents \citep{achiam2023gpt, bubeck2023sparks, wan2023kelly}.
Within auctions it complements three strands: work that treats LLMs as market participants in their own right and studies their collusion, information use, and trainability \citep{fish2024algorithmic, chen2023put, infobid2025, lu2025claude, tsuchihashi2023homo, auctionnet2024}; work that brings LLM bid distributions closer to human ones by adjusting them on a small human sample \citep{corrigan2025usefulness}, which asks a question downstream of ours; and work that designs auctions for LLM-powered platforms \citep{duetting2024mechanism, ramage2024adauctions}.
Markets populated by AI agents themselves are an important adjacent agenda in which LLM behavior is designer-relevant in its own right; our question is when LLM-generated behavior can inform design for human participants.
We do not learn mechanisms, as automated mechanism design does \citep{dutting2024optimal, curry102automated, wang2024gemnet, feng2018deep, wang2025bundleflow, ravindranath2023data}; we provide a benchmark for the behavioral inputs such pipelines would need, and a diagnostic for which of those inputs an LLM panel can supply.
A broader literature connects LLMs with game-theoretic modeling of strategic environments \citep{daskalakis2024charting, huang2025accelerated, soumalias2025llm, jiang2025incentive, wang2026llm}.}

\rev{We release the simulation framework as an open-source benchmark that runs auction experiments with any number of bidders, arbitrary prompting, and many auction formats.\footnote{Code and data: \url{https://github.com/KeHang-Zhu/llm-auction}.}
The experiments reported here comprise more than 1,000 auctions with more than 5,000 GPT-4o bidder participants at a total cost below \$400, plus the four-model panel.
Section~\ref{sec:general_method} describes the reconstruction of human benchmarks, the metric and the three tiers, the simulation protocol, and the model panel.
Section~\ref{sec:benchmark_experimental} reports the laboratory benchmarks tier by tier.
Section~\ref{session:ebay} applies the framework to the clock-framing intervention and to closing rules in an eBay-style marketplace.
Section~\ref{sec:discussion} draws the implications for using LLM panels in mechanism design, and Section~\ref{sec:conclusion} concludes.}

\section{Methods}\label{sec:general_method}
\subsection{Human Experiment Data Collection}

\rev{The experimental auctions literature is rich but fragmented: each study reports its own summary statistics (overbidding frequencies, bid-on-value regressions, mean absolute deviations from the dominant strategy) for its own environment (number of bidders, value support, rounds, interface). To compare LLM behavior with this evidence \emph{across} mechanisms, we treat the published moments as the data. For each source we fit a simple model of individual bidding to the reported moments, simulate bids from the fitted model, and evaluate the same statistic that we compute for LLM bids: the scaled mean absolute deviation from theory defined in Section~\ref{sec:method}. This turns heterogeneous legacy results into one cross-mechanism benchmark. The reconstruction is not a substitute for micro data, so we carry its uncertainty forward as reconstruction bands (Appendix~\ref{app:bands}) and use the reconstructed benchmarks only for orderings, directions, and comparative statics, never for level comparisons or hypothesis tests.}

\paragraph{Reconstructing human bid distributions via moment matching.}
Because micro-level bid data are rarely available from published experiments, we reconstruct synthetic human bid distributions by fitting a mixture model to aggregate statistics
reported in each source. We model bids as:
\[
b = b^\star_m(v) \cdot r
\]
where $b^\star_m(v)$ is the equilibrium bid and $r$ is a bid-to-equilibrium ratio drawn from a three-component mixture:
\begin{align*}
\text{With probability } p_{\text{eq}}: \quad & r \sim \mathcal{N}(1, \sigma^2) \\
\text{With probability } p_{\text{over}}: \quad & r \sim 1 + \text{Exp}(\lambda) \\
\text{With probability } p_{\text{under}}: \quad & r \sim \text{Uniform}(1-\delta, 1)
\end{align*}
This structure maps directly to how experimental papers categorize behavior: equilibrium bidders ($r \approx 1$), overbidders ($r > 1$), and underbidders ($r < 1$).\rev{\footnote{Quantal Response Equilibrium predicts symmetric deviations around equilibrium, contradicting the well-documented asymmetric overbidding in second-price auctions \citep{kagel1993independent}. The mixture model directly maps to how experimental papers categorize and report behavior.}}

\paragraph{Source papers and equilibrium benchmarks.}
We draw on four sources: \citet{li2017obviously} and \citet{breitmoser2022obviousness} for second-price and ascending clock auctions; \citet{kagel1993independent} for first-price,
second-price, and third-price sealed-bid auctions with $n \in \{5, 10\}$ bidders; and \citet{gonczarowski2023strategyproofness} for second-price auctions comparing mechanism
descriptions. For each format, we use the risk-neutral Bayes-Nash equilibrium:
\begin{align*}
b^\star_{\text{SPSB}}(v) &= v & &\text{(second-price, ascending clock)} \\
b^\star_{\text{FPSB}}(v) &= \tfrac{n-1}{n} v & &\text{(first-price)} \\
b^\star_{\text{TPSB}}(v) &= \tfrac{n-1}{n-2} v & &\text{(third-price)}
\end{align*}

\paragraph{Parameter identification.}
Mixture weights $(p_{\text{eq}}, p_{\text{over}}, p_{\text{under}})$ are set directly from reported rates. For example, \citet{li2017obviously} reports ${\sim}50\%$ dominant-strategy
play and ${\sim}40\%$ overbidding, yielding $p_{\text{eq}} = 0.50$, $p_{\text{over}} = 0.40$, $p_{\text{under}} = 0.10$.

The overbid scale $\lambda$ is chosen to match a secondary moment:
\begin{itemize}
  \item For \citet{li2017obviously} and \citet{breitmoser2022obviousness}: the mean bid-to-value ratio (e.g., $\mathbb{E}[b/v] \approx 1.08$).
  \item For \citet{kagel1993independent}: the regression coefficient $\beta$ from bid regressions.
  \item For \citet{gonczarowski2023strategyproofness}: mean absolute deviation (MAD $\approx$ \$0.51--0.55).
\end{itemize}
The equilibrium noise $\sigma$ is derived from regression $R^2$ where reported, via $\sigma^2 = \beta^2 \text{Var}(v)(1-R^2)/R^2$. The underbid bound $\delta$ is set consistently across
papers; we fix it at a common value because papers do not report underbid tail shapes separately.\rev{\footnote{Some reported statistics are difficult to reconcile within any parametric model. For instance, the regression coefficients and bidding frequencies in \citet{kagel1993independent} cannot both be matched under Gaussian noise, suggesting heteroskedasticity in the original data. Our approach matches key aggregate patterns while acknowledging this limitation. Comparisons between LLM and human bidding should be interpreted qualitatively.}}

Full parameter values appear in Appendix~\ref{app:mixture}.

\paragraph{\rev{Reconstruction uncertainty.}}
\rev{Beyond Monte Carlo noise at fixed mixture parameters, the fitted parameters themselves are uncertain: reported rates are rounded, regression coefficients carry standard errors, and the underbid bound $\delta$ is fixed by convention. Appendix~\ref{app:bands} reports a parametric bootstrap over the admissible ranges of every mixture parameter ($B=2000$). The resulting 95\% reconstruction bands are shown as whiskers on the human bars in Figure~\ref{fig:comparison} and in brackets in Table~\ref{tab:model_comparison}. The two load-bearing human orderings survive the bands: first-price deviations exceed second-price deviations in 100\% of re-draws, and the ascending clock improves on sealed second-price bidding in 100\% (open clock) and 92.6\% (closed clock) of re-draws.}

\subsection{\texorpdfstring{\rev{Unified Deviation Metric across Auctions}}{Unified Deviation Metric across Auctions}}\label{sec:method}

\paragraph{Unified deviation metric (applied to all bids).}
Consider an auction \emph{environment} and \emph{format} indexed by $m$.
The index $m$ collects all primitives needed to define the theoretical benchmark, including the number of bidders $N$, the value (or signal) distribution, the information structure (independent private value/common value), and the mechanism rules (e.g., first-price vs.\ second-price, sealed-bid vs.\ clock).
Let $\mathcal I_i$ denote bidder $i$'s information at the moment of bidding under environment $m$ (e.g., $\mathcal I_i=v_i$ in private values, or $\mathcal I_i=s_i$ in common values).
Let $b_{i}$ denote an \emph{observed submitted bid} by bidder $i$ (we apply the definition to \emph{all bids}, not only winning bids), and let $b_m^\star(\mathcal I_i)$ denote the \emph{theoretical benchmark bid} under $m$ given information $\mathcal I_i$ (Bayes--Nash equilibrium in non-strategy-proof formats; truthful/value bidding in strategy-proof formats).

\noindent We define the absolute bid error for a bid with information $\mathcal I_i$ as
\[
d_m(\mathcal I_i)\;\equiv\;\bigl|b_i - b_m^\star(\mathcal I_i)\bigr|.
\]
To compare deviations across formats on a common scale, we normalize by the expected benchmark bid level,
\[
\mu_m^\star \;\equiv\; \mathbb E_m\!\left[b_m^\star(\mathcal I)\right],
\]
where $\mathbb E_m[\cdot]$ denotes expectation with respect to the distribution of the information object $\mathcal I$ induced by environment $m$ (i.e., the value/signal distribution and any randomization in the experimental design).
We then define the \emph{scaled mean absolute deviation} (SMAD) for auction format $m$ as
\[
\Delta_m \;\equiv\; 100 \cdot \frac{\mathbb E_m\!\left[d_m(\mathcal I)\right]}{\mu_m^\star}
\;=\; 100 \cdot \frac{\mathbb E_m\!\left[\left|b-b_m^\star(\mathcal I)\right|\right]}{\mathbb E_m\!\left[b_m^\star(\mathcal I)\right]}.
\]
This metric is interpretable as a percentage of the expected benchmark bid: $\Delta_m=10$ means that, on average, bids deviate from the benchmark by an amount equal to $10\%$ of the expected benchmark bid.

\paragraph{\rev{What the human literature says about the ordering of SMAD.}}
\rev{The value of a common scale is that the experimental literature's conclusions about \emph{relative} difficulty become testable statements about SMAD. Three such conclusions anchor our benchmark. First, implementation matters even when incentives do not: the same strategy-proof second-price rule is played more truthfully as an ascending clock than as a sealed bid \citep{li2017obviously, breitmoser2022obviousness}. In the reconstructed anchors, SMAD is $3.5\%$ in the open clock and $5.8\%$ in the closed clock against $9.3\%$ in the sealed-bid version of the same affiliated-values environment. Second, a dominant strategy is easier to play than an equilibrium that requires bid shading: second-price bidders deviate far less than first-price bidders in the same independent-private-values environment ($5.6\%$ against $24.8\%$, a $4.4\times$ gap; \citealp{kagel1993independent}). Third, common-value environments, in which a bidder must condition on the adverse-selection content of winning, are harder than private-value environments with the same payment rule ($47.6\%$ against $24.8\%$ under first-price rules and $18.2\%$ against $5.6\%$ under second-price rules), and within common values the first-price winner's curse is more severe than the second-price curse \citep{kagel1986winner, levin1996revenue}. Ordered by human SMAD, the seven formats therefore run from first-price common value ($47.6\%$) through first-price IPV ($24.8\%$), second-price common value ($18.2\%$), sealed second-price APV ($9.3\%$), the closed clock ($5.8\%$), and second-price IPV ($5.6\%$) to the open clock ($3.5\%$). It is this ordering, and the signs of the three design contrasts behind it, that we test LLM bidders against; the magnitudes themselves carry the reconstruction uncertainty discussed above.}

Given a sample of bids $\{(b_k,\mathcal I_k)\}_{k=1}^n$ from format $m$, the plug-in estimator is
\[
\widehat{\Delta}_m
\;=\;
100 \cdot
\frac{\frac{1}{n}\sum_{k=1}^{n}\left|b_k-b_m^\star(\mathcal I_k)\right|}
{\widehat{\mu}_m^\star},
\qquad
\widehat{\mu}_m^\star \equiv \frac{1}{n}\sum_{k=1}^{n} b_m^\star(\mathcal I_k).
\]
When we report a standard error for the (unscaled) mean absolute deviation $\frac{1}{n}\sum_k |b_k-b_m^\star(\mathcal I_k)|$, we propagate it to $\widehat{\Delta}_m$ by scaling with $\widehat{\mu}_m^\star$. We construct $95\%$ confidence intervals using a normal approximation when an analytic standard error is available.

\paragraph{Deterministic benchmarks.}
SMAD requires a single-valued benchmark mapping $\mathcal I \mapsto b_m^\star(\mathcal I)$.
Accordingly, we restrict attention to auction formats where the theoretical prediction is deterministic given the bidder’s information.

\subsection{LLM Data Generation}
\label{sec:data_generation}

We generate synthetic auction data by simulating auctions in which $n$ large language model (LLM) agents bid against one another under fully specified information structures and auction rules. Each experimental \emph{setting} is defined by (i) a \emph{value model} (how bidder values/signals are drawn), (ii) a \emph{mechanism environment} (the action space, allocation and payment rules, and the information revealed during play), and (iii) a \emph{prompt framing} (how the mechanism is described to the bidder). For each setting, we repeat the simulation $K$ times with independent random seeds, producing a dataset of realized values (or signals), bidder actions, allocations, prices, profits, and free-text plans. Here, ``actions'' refers to: a numeric bid in sealed-bid formats, a \texttt{stay}/\texttt{exit} decision at each clock tick in ascending auctions, and an \texttt{increase}/\texttt{hold} decision over a bidder's maximum bid in the eBay proxy-bidding environment.

\paragraph{Single-auction baseline and learning ablation.}
Our main benchmark experiments use \emph{single-auction} simulations: each repetition $k$ is an independent auction instance with freshly drawn values/signals, and bidders do not receive feedback across instances. We adopt this design to minimize reliance on cross-instance learning and long-term adaptation, which remains unreliable for language models \citep{dong2024survey}. Importantly, this choice is distinct from whether a mechanism is \emph{iterative} within an auction (e.g., clock auctions): iterative mechanisms still unfold over multiple within-auction steps (clock ticks), but we do not run repeated auctions with feedback by default. In Appendix~\ref{app:learning}, we run a repeated-auction ablation for sealed-bid auctions in which bidders observe outcomes from prior auction instances; consistent with the motivation above, we find no meaningful learning effects across repetitions.

\paragraph{Value models and numerical parameters.}
We implement three canonical information structures. All parameters (support ranges, bidder count $n$, and bid/price increments) are specified in configuration files and held fixed within a setting. In sealed-bid auctions, bids are restricted to a discrete grid with increment $0.1$; in ascending clock auctions, the clock price increases by increment $0.5$ per tick.

\begin{itemize}
    \item \textbf{Independent Private Values (IPV).} Each bidder's value is the sum of an independent private component and (optionally) a common component. In our baseline IPV parameterization, the private component from $[0,49]$, yielding
    \[
          \epsilon_{it} \sim \mathrm{Unif}[0,49], \qquad v_{it}=\epsilon_{it}.
    \]
    We discretize bids with increment $0.1$.

    \item \textbf{Affiliated Private Values (APV).} Values share a common component plus an idiosyncratic private shock. Our APV parameterization sets
    \[
        c_t \sim \mathrm{Unif}[0,29], \qquad \epsilon_{it} \sim \mathrm{Unif}[0,20], \qquad v_{it}=c_t+\epsilon_{it},
    \]
    again with bid increment $0.1$. This induces affiliation: higher realized values for one bidder make higher values for others more likely.

    \item \textbf{Common Value (CV).} The item has an underlying common value $V_t$ shared by all bidders, while each bidder observes a noisy private signal. In our CV parameterization, the common value is drawn from $[20,29]$ and each bidder observes an idiosyncratically perturbed signal,
    \[
        V_t \sim \mathrm{Unif}[20,29], \qquad \eta_{it} \sim \mathrm{Unif}[-20,20], \qquad s_{it}=V_t+\eta_{it},
    \]
    with bid increment $0.1$. Bidder $i$ conditions on $s_{it}$ when bidding, while realized payoffs depend on $V_t$, generating the winner's-curse selection effect.
\end{itemize}
In our baseline benchmark, $T=1$ so each repetition corresponds to a single auction instance.

\paragraph{\rev{Bidder instructions.}}
\rev{Every simulated bidder is assigned a value (or signal) and receives the same instruction to maximize expected profit. This choice follows the design of the human experiments we benchmark against. Laboratory auctions are incentivized: subjects' earnings rise with their auction payoffs, and the instructions encourage them to earn as much as they can \citep{smith1976experimental, smith1982microeconomic}. The experiments of \citet{kagel1993independent}, \citet{li2017obviously}, \citet{breitmoser2022obviousness}, and \citet{gonczarowski2023strategyproofness} all induce values in this way and pay subjects accordingly. Within each model and setting, bidders share the same objective and differ in their assigned values or signals. Human subjects additionally differ in preferences, experience, and cognition, even under the same induced-value instructions; we do not assign demographic or personality descriptions, which is one reason to expect magnitudes to differ (Section~\ref{sec:discussion}). In the one-shot baseline the objective instruction reads: ``You have only one opportunity to bid. Your top priority is to place a bid that maximizes your expected profit.''; a multi-round variant is used only in the learning ablation (Appendix~\ref{app:learning}).}
%
%

\subsubsection{Canonical laboratory mechanisms.}
To mirror standard laboratory benchmarks, we implement \textit{sealed-bid} auctions and \textit{ascending clock} auctions under the value models above.

\paragraph{Sealed-bid auctions.}
In sealed-bid auctions, each bidder submits a single bid $b_{it}$ simultaneously. Let $b_{(1)t}\ge b_{(2)t}\ge b_{(3)t}$ denote the order statistics and let $i^\star_t$ be the highest bidder. The winner pays a price $p_t$ determined by the mechanism:
\begin{itemize}
    \item \textbf{First-price sealed-bid (FP).} $i^\star_t$ wins and pays $p_t=b_{(1)t}$.
    \item \textbf{Second-price sealed-bid (SP).} $i^\star_t$ wins and pays $p_t=b_{(2)t}$.
\end{itemize}
Profits are computed as $\pi_{it}=v_{it}-p_t$ for winners (or $\pi_{it}=V_t-p_t$ in CV) and $0$ for losers.

\paragraph{Ascending clock auctions.}
In an ascending clock auction, the price starts at 0 and increases by a fixed increment $\Delta_{\text{clock}}=0.5$ at each clock tick. At each tick, each bidder chooses \emph{stay} or \emph{exit}. The last remaining bidder wins. We implement the standard second-price logic in the dynamic interface: the winner pays the exit price of the second-to-last bidder.

\paragraph{\rev{Clock-framing intervention (mechanism held fixed).}}
To isolate the effect of description while holding auction design constant, we implement \rev{an alternative framing} of the second-price sealed-bid auction:
\begin{itemize}
    \item \textbf{Clock framing.} Bidders are told to treat their sealed bid as an automatic exit threshold in a simulated ascending clock: conceptually, the clock rises until bidders ``drop out'' at their thresholds, and the winner pays the second-highest threshold.
\end{itemize}
\rev{The framing implements} the same allocation and payment rule as the second-price sealed-bid auction; only the prompt language differs. The precise prompts are listed in Appendix~\ref{app:prompt}.

\rev{The eBay-style proxy-bidding environment used in Section~\ref{session:ebay} is a separate, marketplace-inspired environment; its rules, treatments, and discrete-time implementation are described in Appendix~\ref{app:ebay_env}.}

\subsubsection{Simulation protocol.}
Each experiment proceeds as follows, repeated for $K$ independent repetitions:
\begin{enumerate}
    \item \textbf{Value realization.} For each repetition $k$, we draw values (IPV/APV) or signals (CV) for all bidders under a fixed random seed. In our baseline benchmark experiments, each repetition corresponds to a single auction instance.
    \item \textbf{Prompted action.} We construct an individualized prompt for each bidder containing (i) the mechanism rules (and framing, if applicable), (ii) the bidder's current value (or signal), and (iii) when applicable, a feedback transcript summarizing prior auction instances (only in the learning ablation). We request a structured response with explicit tags:
\begin{quote}\small
\verb|<PLAN> ... </PLAN>| \\
\verb|<ACTION> ... </ACTION>|
\end{quote}
where \verb|<ACTION>| contains a numeric bid (sealed-bid), a stay/exit decision (clock), or an increase/hold decision over the bidder's maximum bid (eBay).
    \item \textbf{Parallel elicitation and parsing.} We query the LLM for each bidder using separate API calls in parallel, and parse the action from the \verb|<ACTION>| tag using a deterministic regex-based extractor.
    \item \textbf{Outcome computation.} Given the submitted actions, we determine the winner and payments according to the mechanism rules and compute bidder profits.
\end{enumerate}

Unless otherwise noted, our baseline settings use $n=3$ bidders and we repeat each configuration $K$ times with independent seeds to ensure stable distributional estimates.

\subsection{\texorpdfstring{\rev{Three Tiers of Comparison}}{Three Tiers of Comparison}}\label{sec:tiers}

\rev{We compare LLM and human behavior at three tiers.}
\begin{itemize}
  \item \rev{\emph{Magnitudes.} The size of SMAD. Because the human anchors are reconstructions, magnitude comparisons are qualitative, and we ask whether a model's SMAD falls within the human reconstruction band.}
  \item \rev{\emph{Directions.} After removing bids within $\pm 2\%$ of the benchmark, $|b-b^\star| \le 0.02\, b^\star$, we classify the remaining bids as underbids or overbids and compare the shares. Human direction shares use the source studies' reported categories in the truthful formats and the reconstruction in first-price auctions.}
  \item \rev{\emph{Orderings.} We rank formats by SMAD and compute Kendall's $\tau_b$ between the human ranking and each model's ranking over the five private-value formats that share a common benchmark scale (FPSB IPV, SPSB IPV, SP-APV, AC, AC-B), with uncertainty from two layers: resampling LLM bids within each cell, and re-drawing the human reconstruction within its admissible ranges (Appendix~\ref{app:concordance}). For design contrasts we ask whether the \emph{sign} of the human effect (first-price harder than second-price; clock easier than sealed bid) is reproduced.}
\end{itemize}
\rev{Effect sizes and error directions matter for design objectives such as revenue and efficiency. Ordinal agreement can inform comparisons between mechanisms, and it is the tier at which we claim LLM bidders are informative.}

\subsection{\texorpdfstring{\rev{Model Panel}}{Model Panel}}\label{sec:models}

\rev{We evaluate five LLMs under identical prompts, in three classes. Three are large models without extended test-time reasoning, which we call the non-reasoning large model family: OpenAI's gpt-4o (2024-08-06), Anthropic's claude-3-5-haiku (2024-10-22), and Google's gemini-2.0-flash (2025-02-05). One is a reasoning model that allocates extended test-time computation to each decision, gpt-5-mini (2025-08-07). One is a small open-weights model, Meta-Llama-3-8B-Instruct. Each model is run in separate auctions with bidders using that model; the panel compares results across models rather than mixing models within an auction. Section~\ref{sec:benchmark_experimental} focuses on the three non-reasoning large models and reports the reasoning and small models as comparisons. GPT-4o is in addition the reference model for the deeper distributional diagnostics (temperature ablation, repeated rounds, and the eBay-style environment of Section~\ref{session:ebay}) and therefore has the largest cells. For non-reasoning models we set the temperature to $0.5$, the most common setting in LLM simulation \citep{horton2023large, Horton2024EDSL}; for gpt-5-mini the API ignores temperature. For GPT-4o we also ablate $\text{temp}\in\{0.1, 1.0\}$ (Appendix~\ref{app:ablation}).}
%
%
%
%
%
%
%
%
%
%

\section{Benchmark Experimental Auction Data}\label{sec:benchmark_experimental}

\begin{figure}[tbp]
    \centering
\includegraphics[width=0.85\linewidth]{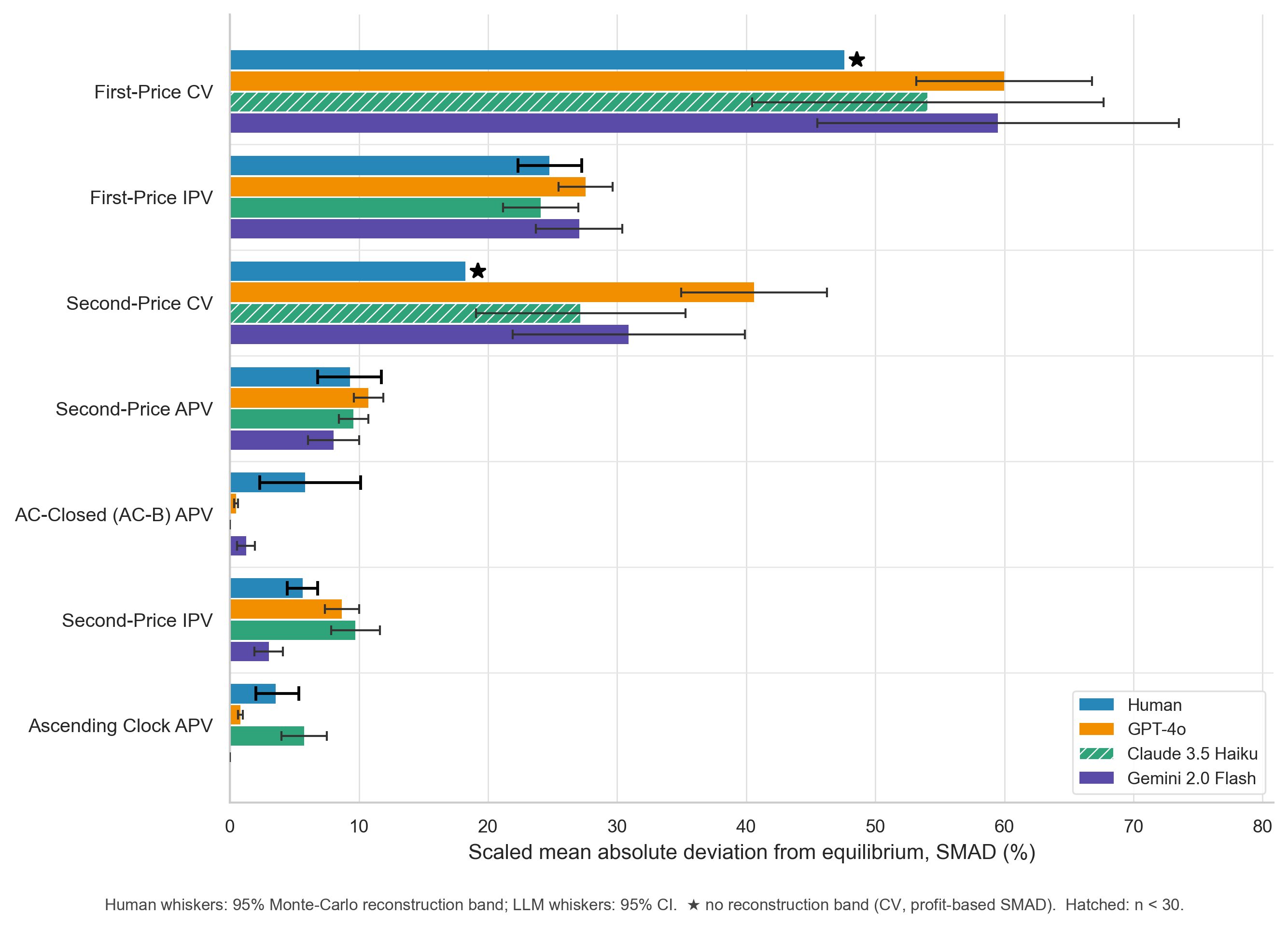}
    \caption{\rev{\textbf{Human vs.\ LLM bidding behavior across seven auction formats, non-reasoning large models.} Scaled mean absolute deviation (SMAD) from the theoretical benchmark for the reconstructed human data (blue; black whiskers are 95\% reconstruction bands from the parametric bootstrap of Appendix~\ref{app:bands}; common-value anchors have no band) and for the three non-reasoning large models GPT-4o, Claude~3.5 Haiku, and Gemini~2.0 Flash (95\% confidence intervals), with $N=3$ bidders. Hatched bars have fewer than 30 usable bids. Formats are ordered by human SMAD. The reasoning model GPT-5-mini and the small model Llama-3-8B are reported in Table~\ref{tab:model_comparison} and in Figure~\ref{fig:model_ablation}.}
    \label{fig:comparison}}
\end{figure}

\rev{Figure~\ref{fig:comparison} and Table~\ref{tab:model_comparison} report SMAD for the seven laboratory formats, ordered by human SMAD from hardest to easiest, for the reconstructed human data and the five models. We read them tier by tier (Section~\ref{sec:tiers}): magnitudes, directions, orderings.}

\begin{table}[htbp]
\centering
\small
\rev{\resizebox{\linewidth}{!}{\begin{tabular}{lcccccc}
\toprule
Format & Human & GPT-4o & Claude 3.5 Haiku & Gemini 2.0 Flash & GPT-5-mini & Llama-3-8B \\
\midrule
First-Price CV & 47.59$^{\ast}$ & 59.99 {\scriptsize (96)} & 54.07 {\scriptsize (29)}$^{\dagger}$ & 59.50 {\scriptsize (30)} & 30.22 {\scriptsize (30)} & 53.68 {\scriptsize (30)} \\
First-Price IPV & 24.76 [22.30, 27.25] & 27.55 {\scriptsize (285)} & 24.09 {\scriptsize (90)} & 27.05 {\scriptsize (90)} & 0.08 {\scriptsize (87)} & 38.16 {\scriptsize (90)} \\
Second-Price CV & 18.23$^{\ast}$ & 40.60 {\scriptsize (92)} & 27.19 {\scriptsize (29)}$^{\dagger}$ & 30.89 {\scriptsize (30)} & 14.80 {\scriptsize (30)} & 32.27 {\scriptsize (30)} \\
Second-Price APV & 9.31 [6.78, 11.74] & 10.73 {\scriptsize (291)} & 9.58 {\scriptsize (90)} & 8.02 {\scriptsize (87)} & 1.36 {\scriptsize (87)} & 53.08 {\scriptsize (90)} \\
AC-Closed (AC-B) APV & 5.83 [2.30, 10.11] & 0.47 {\scriptsize (114)} & 0.00 {\scriptsize (2)}$^{\dagger}$ & 1.26 {\scriptsize (46)} & DS$^{\ddagger}$ & 67.00 {\scriptsize (2)}$^{\dagger}$ \\
Second-Price IPV & 5.65 [4.46, 6.79] & 8.69 {\scriptsize (270)} & 9.72 {\scriptsize (90)} & 3.01 {\scriptsize (87)} & 0.20 {\scriptsize (87)} & 54.89 {\scriptsize (90)} \\
Ascending Clock APV & 3.54 [2.00, 5.36] & 0.81 {\scriptsize (106)} & 5.74 {\scriptsize (46)} & 0.00 {\scriptsize (2)}$^{\dagger}$ & DS$^{\ddagger}$ & 104.10 {\scriptsize (8)}$^{\dagger}$ \\
\bottomrule
\end{tabular}
}}
\caption{\rev{\textbf{SMAD (\%) by auction format and model.} Human anchors are moment-matched reconstructions; brackets give 95\% reconstruction bands ($^{\ast}$: common-value anchors are profit-based moments and carry no band). LLM cells report SMAD with the number of usable bids in parentheses; $^{\dagger}$ marks cells with fewer than 30 bids. DS$^{\ddagger}$ marks the two ascending-clock cells of GPT-5-mini, in which the model states the dominant strategy outright and does not complete the tick-by-tick clock procedure in any run, so no bids are recorded; these cells enter no statistic. All models run at temperature $0.5$ (ignored by GPT-5-mini). We do not report significance tests of LLM cells against the reconstructed anchors, which are reconstructions of published moments rather than samples.}}
\label{tab:model_comparison}
\end{table}

\subsection{\texorpdfstring{\rev{Tier 1: Magnitudes do not match}}{Tier 1: Magnitudes do not match}}\label{sec:tier_magnitudes}

\rev{Table~\ref{tab:model_comparison} compares SMAD magnitudes. Against the human reconstruction bands of Appendix~\ref{app:bands}, the reference model GPT-4o falls inside the band in one of the five private-value formats (SP-APV, $10.7\%$ vs.\ $[6.8, 11.7]$); it deviates more than humans in the sealed-bid formats ($27.6\%$ vs.\ $24.8\%$ in first-price, $8.7\%$ vs.\ $5.6\%$ in second-price) and far less in the clock formats ($0.8\%$ vs.\ $3.5\%$ in the open clock, $0.5\%$ vs.\ $5.8\%$ in the closed clock). In the common-value formats its deviations are $26\%$ and $123\%$ larger than the human anchors. The other models do no better: Claude~3.5 Haiku and Gemini~2.0 Flash fall inside the human band in two of five private-value formats each, GPT-5-mini in none (its SMAD is below $1.4\%$ in every private-value format, essentially the theoretical benchmark), and Llama-3-8B in none (its SMAD exceeds $38\%$ everywhere). Even where a single magnitude coincides, as with Claude's first-price SMAD of $24.1\%$ against the human $24.8\%$, the same model is off by a factor of two in the adjacent second-price format. We conclude that LLM simulations should not be read for the magnitude of human deviations, and the rest of this section therefore turns to the two qualitative tiers.}

\subsection{\texorpdfstring{\rev{Tier 2: Directions do not match}}{Tier 2: Directions do not match}}\label{sec:tier_directions}

\begin{figure}[tbp]
    \centering
    \includegraphics[height=0.78\textheight,keepaspectratio]{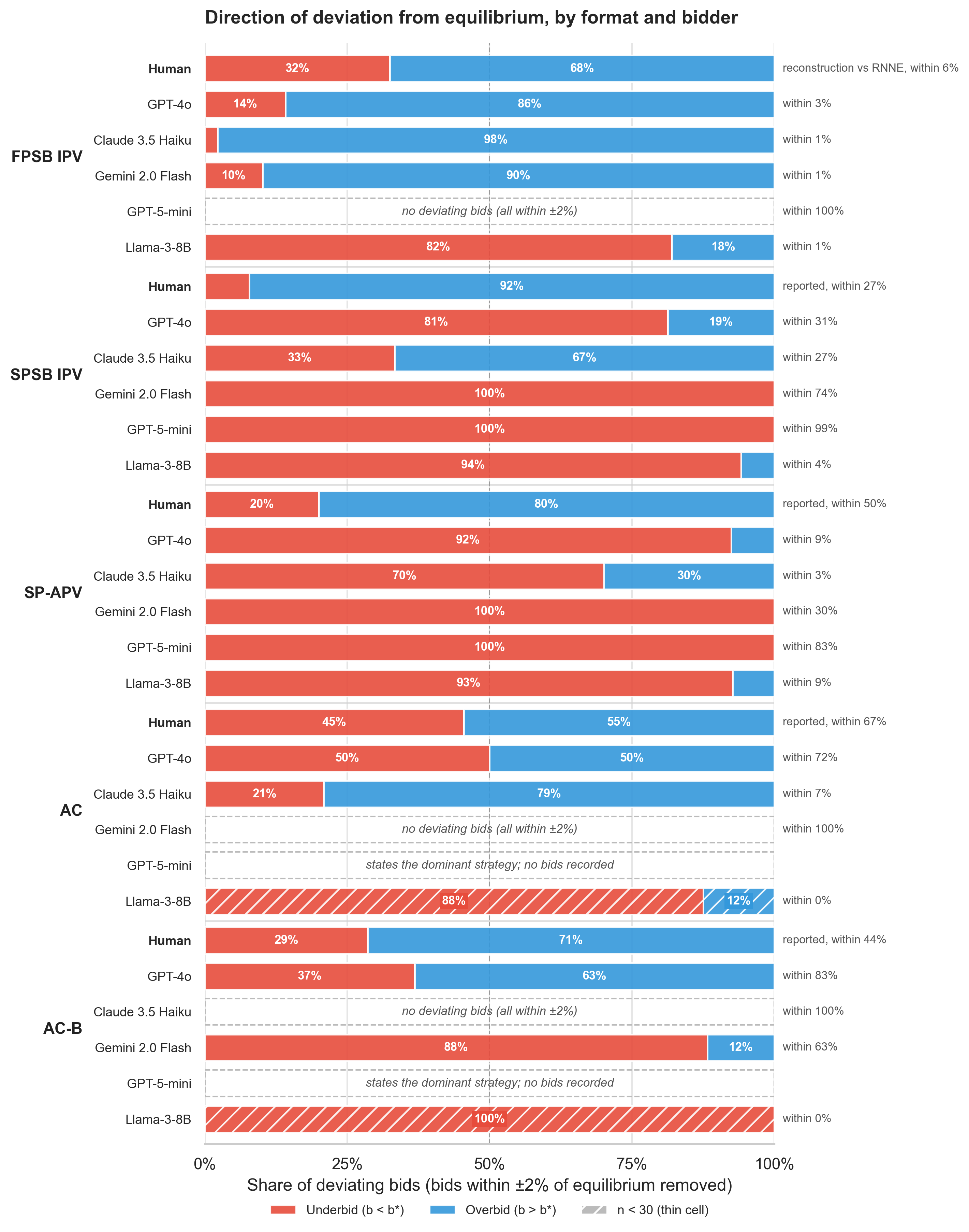}
    \caption{\rev{\textbf{Direction of deviating bids by model.} Shares of underbids (red) and overbids (blue) after removing bids within $\pm2\%$ of the theoretical benchmark, for human bidders and each model, by auction format. Human shares are the frequencies reported by the source studies relative to value for the truthful formats, and the moment-matched reconstruction relative to the equilibrium bid for first-price. Hatched bars have fewer than 30 usable bids; in the clock formats GPT-5-mini states the dominant strategy and records no bids.}
    \label{fig:overbidding}}
\end{figure}

\rev{Figure~\ref{fig:overbidding} compares the direction of bid errors. LLM shares exclude bids within $\pm 2\%$ of the theoretical benchmark. Human shares use the source studies' reported categories in the truthful formats and the reconstructed bids in first-price auctions, so their tolerance convention need not match the LLM convention.}

\rev{In first-price IPV auctions, humans and the non-reasoning large models all predominantly overbid relative to the equilibrium shading rule. Overbids account for $67.5\%$ of human deviating bids, after removing $6.4\%$ near-equilibrium bids, and for $85.9\%$, $97.8\%$, and $89.9\%$ of the deviating bids of GPT-4o, Claude~3.5 Haiku, and Gemini~2.0 Flash, after removing $3.2\%$, $1.1\%$, and $1.1\%$. The small model underbids ($82\%$ of Llama-3-8B's deviating bids), and the reasoning model has no deviating bids: every GPT-5-mini bid lies within the tolerance band around the shading rule. Second-price IPV reverses the agreement. \citet{kagel1993independent} report $67\%$ of human bids above value, $27\%$ at value, and $6\%$ below, implying that about $92\%$ of non-truthful bids are overbids. Among GPT-4o's deviating bids, $81.3\%$ are underbids, and $30.7\%$ of its bids lie within the tolerance band; Gemini~2.0 Flash underbids in all of its deviating bids, whereas Claude~3.5 Haiku overbids in roughly two thirds of its deviating bids, the one member of the family that errs in the human direction. The small model also predominantly underbids ($94\%$), and the reasoning model has a single deviating bid.}

\rev{In sealed-bid SP-APV, the aggregate deviations are similar for humans and GPT-4o ($9.3\%$ and $10.7\%$), but the errors have opposite directions: $80\%$ of human deviating bids are overbids, while $92.4\%$ of GPT-4o's are underbids, as are $70\%$ of Claude~3.5 Haiku's and all of Gemini~2.0 Flash's. All five models predominantly underbid in this setting. The open clock gives a more similar directional split for humans and GPT-4o ($45/55$ versus $50/50$ under/over), although their SMAD magnitudes still differ, and the closed clock behaves alike ($29/71$ versus $37/63$). Directional agreement is thus distinct from agreement in either the average error or the full bid distribution.}

\subsection{\texorpdfstring{\rev{Tier 3: Orderings match for the non-reasoning large models}}{Tier 3: Orderings match for the non-reasoning large models}}\label{sec:tier_orderings}

\rev{The ordinal tier asks whether the mechanisms that are harder for humans are also harder for LLM bidders, and whether the design contrasts that move human behavior move LLM behavior in the same direction. We report the ranking as a whole and the scorecard first, then the three design contrasts on which the human literature is clearest, narrating each by model class: the non-reasoning large models, the reasoning model, and the small model.}

\paragraph{\rev{Rank concordance.}}
\rev{Table~\ref{tab:concordance} reports Kendall's $\tau_b$ between the human SMAD ranking and each model's ranking over the five private-value formats, with uncertainty from resampling LLM bids within each cell and from re-drawing the human reconstruction. The non-reasoning large models reproduce the human ranking. For GPT-4o, $\tau_b = 0.60$ over all five formats (8 concordant and 2 discordant pairs, both involving the closed clock); the estimate is unchanged under bid-level resampling and remains positive in $100\%$ of joint draws that also re-draw the human reconstruction (joint 95\% CI $[0.40, 1.00]$), with first-price ranked hardest on both sides in every draw. For Claude~3.5 Haiku and Gemini~2.0 Flash, $\tau_b = 0.67$ over the four formats with adequate cells, again positive in essentially every joint draw. GPT-5-mini's deviations are only $0.1\%$--$1.4\%$ in the observed private-value formats; its $\tau_b$ of $-0.33$ ranks small differences among near-equilibrium bids, unlike the substantial deviations observed in humans. The small model inverts the ranking: Llama-3-8B's $\tau_b$ is $-1.00$, with deviations above $38\%$ in every format. The human ordering is thus recovered by the non-reasoning large models, not by LLMs as such, and for GPT-4o it is also robust to the sampling temperature (Appendix~\ref{app:temperature}).}

\begin{table}[htbp]
\centering
\small
\rev{\resizebox{\linewidth}{!}{\begin{tabular}{llrlrlr}
\toprule
Model & Formats used ($n\geq 30$) & $\tau_b$ (all usable) & Joint 95\% CI & $P(\tau_b>0)$ & $\tau_b$ (3 sealed) & FPSB$>$SPSB \\
\midrule
GPT-4o & all five (5) & $+0.60$ & $[+0.40, +1.00]$ & 100\% & $+1.00$ & yes (100\%) \\
Claude 3.5 Haiku & 3 sealed + AC (4) & $+0.67$ & $[+0.67, +1.00]$ & 100\% & $+0.33$ & yes (100\%) \\
Gemini 2.0 Flash & 3 sealed + AC-B (4) & $+0.67$ & $[+0.33, +1.00]$ & 100\% & $+1.00$ & yes (100\%) \\
GPT-5-mini & 3 sealed (3) & $-0.33$ & $[-0.33, +0.33]$ & 36\% & $-0.33$ & no (36\%) \\
Llama-3-8B & 3 sealed (3) & $-1.00$ & $[-1.00, -0.33]$ & 0\% & $-1.00$ & no (0\%) \\
\bottomrule
\end{tabular}
}}
\caption{\rev{Kendall's $\tau_b$ between the human and each model's SMAD ranking over the private-value formats, with joint 95\% bootstrap intervals over LLM bid resampling and human reconstruction re-draws ($B=2000$). GPT-5-mini's clock cells record no bids (Table~\ref{tab:model_comparison}) and are excluded.}}
\label{tab:concordance}
\end{table}

\paragraph{\rev{Scorecard.}}
\rev{Table~\ref{tab:scorecard} collects the three tiers in one place. Each row is a feature of human behavior established in the experimental literature; each column records whether a model reproduces it. The magnitude rows show that no model falls within the human reconstruction band in more than two of the five private-value formats. The direction rows show that the human direction of second-price errors is reproduced by one model only. The ordering rows show that every ordering with a common benchmark is reproduced by the non-reasoning large models, and that the common-value ordering is reproduced by all five; the two models that fail the other orderings differ from humans in opposite ways: GPT-5-mini has very small private-value errors, while Llama-3-8B has large errors whose ordering is opposite to humans'. For a designer, the table says what LLM simulation can be used for in this domain: to screen which of two designs is behaviorally harder, not to predict how far, or in which direction, human bidders will deviate.}

\begin{table}[htbp]
\centering
\small
\rev{\resizebox{\linewidth}{!}{
\begin{tabular}{lccccccc}
\toprule
Human regularity & Human & GPT-4o & Claude 3.5 Haiku & Gemini 2.0 Flash & GPT-5-mini & Llama-3-8B & Non-reasoning large models \\
\midrule
\multicolumn{8}{l}{\emph{Tier 1: magnitude within the human reconstruction band}} \\
FPSB IPV & [22.3, 27.3] & $\times$ & \checkmark & \checkmark & $\times$ & $\times$ & 2/3 \\
SPSB IPV & [4.5, 6.8] & $\times$ & $\times$ & $\times$ & $\times$ & $\times$ & 0/3 \\
Sealed SP-APV & [6.8, 11.7] & \checkmark & \checkmark & \checkmark & $\times$ & $\times$ & 3/3 \\
Open clock, AC (APV) & [2.0, 5.4] & $\times$ & $\times$ & n/a & DS$^{\ddagger}$ & n/a & 0/2 \\
Closed clock, AC-B (APV) & [2.3, 10.1] & $\times$ & n/a & $\times$ & DS$^{\ddagger}$ & n/a & 0/2 \\
\midrule
\multicolumn{8}{l}{\emph{Tier 2: direction of deviations}} \\
SPSB IPV errors are mostly overbids & \checkmark & $\times$ & \checkmark & $\times$ & $\approx$ & $\times$ & 1/3 \\
Sealed SP-APV errors are mostly overbids & \checkmark & $\times$ & $\times$ & $\times$ & $\times$ & $\times$ & 0/3 \\
\midrule
\multicolumn{8}{l}{\emph{Tier 3: cross-mechanism orderings (SMAD)}} \\
FPSB harder than SPSB (IPV) & \checkmark\ (100\%) & \checkmark & \checkmark & \checkmark & $\approx$ & $\times$ & 3/3 \\
Open clock easier than sealed SP (APV) & \checkmark\ (100\%) & \checkmark & \checkmark$^{a}$ & n/a & DS$^{\ddagger}$ & n/a & 2/2 \\
Closed clock easier than sealed SP (APV) & \checkmark\ (92.6\%) & \checkmark & n/a & \checkmark$^{a}$ & DS$^{\ddagger}$ & n/a & 2/2 \\
FP curse stronger than SP curse (CV) & \checkmark & \checkmark & \checkmark & \checkmark & \checkmark & \checkmark & 3/3 \\
\bottomrule
\end{tabular}
}}
\caption{\rev{\textbf{Scorecard by tier.} Rows are features of human behavior; cells record whether a model reproduces them. Tier 1: SMAD inside (\checkmark) or outside ($\times$) the human 95\% reconstruction band, which the human column reports. Tiers 2 and 3: \checkmark reproduced, $\times$ not reproduced, $\approx$ uninformative because the cells sit at the theoretical benchmark; the human column gives the share of reconstruction re-draws in which the ordering holds where applicable. n/a: fewer than 30 usable bids; $^{a}$: 46 usable bids; DS$^{\ddagger}$: GPT-5-mini states the dominant strategy outright and does not complete the clock procedure, so no clock bids are recorded. The last column counts the non-reasoning large models that reproduce the row, out of those with data.}}
\label{tab:scorecard}
\end{table}

\subsubsection{\texorpdfstring{\rev{Payment rule: first-price is harder than second-price}}{Payment rule: first-price is harder than second-price}}

\rev{Changing the payment rule from first-price to second-price holds the allocation rule fixed but changes what a bidder must compute: a Bayes--Nash shading rule in the first-price auction against a dominant strategy in the second-price auction. Human bidders find the first-price format far harder (SMAD $24.8\%$ against $5.6\%$, a $4.4\times$ gap; \citealp{kagel1993independent}). The non-reasoning large models reproduce the ordering: first-price SMAD exceeds second-price SMAD for GPT-4o ($27.6\%$ vs.\ $8.7\%$), Claude~3.5 Haiku ($24.1\%$ vs.\ $9.7\%$), and Gemini~2.0 Flash ($27.1\%$ vs.\ $3.0\%$), with gaps between $2.5\times$ and $9\times$ around the human $4.4\times$. The reasoning model makes the ordering uninformative: GPT-5-mini bids on the equilibrium shading rule in the first-price auction ($0.1\%$) and at value in the second-price auction ($0.2\%$), leaving nothing to order. The small model inverts it: Llama-3-8B deviates more in the second-price auction ($54.9\%$) than in the first-price auction ($38.2\%$).}

\subsubsection{\texorpdfstring{\rev{Implementation: clock beats sealed bid}}{Implementation: clock beats sealed bid}}

\rev{The second contrast holds the payment rule fixed and changes the implementation: the same second-price logic as a one-shot sealed bid or as an ascending clock in which bidders decide at each price whether to stay or exit. Truthful play is dominant in both, but human bidders play the clock far better, the observation behind obvious strategy-proofness \citep{li2017obviously, breitmoser2022obviousness}: in the affiliated-values environment, human SMAD is $3.5\%$ in the open clock and $5.8\%$ in the closed clock against $9.3\%$ in the sealed bid. The non-reasoning large models reproduce the ordering wherever the clock cells are adequate. GPT-4o's SMAD falls from $10.7\%$ in the sealed bid to $0.8\%$ in the open clock and $0.5\%$ in the closed clock, a larger clock advantage than the human one; Claude~3.5 Haiku's falls from $9.6\%$ to $5.7\%$ in the open clock ($n=46$) and Gemini~2.0 Flash's from $8.0\%$ to $1.3\%$ in the closed clock ($n=46$); the remaining clock cells of these two models have two usable bids each and are not informative. The reasoning model again offers no ordering to test: GPT-5-mini is at the benchmark in the sealed bid ($1.4\%$), and in the clock formats it states the dominant strategy outright and does not complete the tick-by-tick procedure in any run, so no clock bids are recorded (Table~\ref{tab:model_comparison}). The small model does not reproduce the ordering: Llama-3-8B's clock cells have at most eight usable bids, and those bids deviate more than its sealed bids ($104\%$ and $67\%$ against $53\%$). On the human side, the open-clock ordering holds in $100\%$ of reconstruction re-draws and the closed-clock ordering in $92.6\%$, reflecting the wide standard error reported for the closed clock \citep{breitmoser2022obviousness}; we treat the closed-clock ordering as the less secure of the two.}

\subsubsection{\texorpdfstring{\rev{Information structure: common value and competition}}{Information structure: common value and competition}}

\rev{The third contrast changes the information structure. In a common-value auction each bidder observes a noisy signal of a value that is the same for everyone, so winning is bad news: the highest signal is on average too optimistic, and a bidder who does not condition on winning overbids and earns a negative profit, the winner's curse \citep{kagel1986winner}. The human literature establishes that the curse is more severe under first-price than under second-price rules \citep{levin1996revenue}. Because the benchmark for common-value bids is profit-based, we compare SMAD only within the information structure. All three model classes reproduce the ordering. First-price common-value deviations exceed second-price deviations for the non-reasoning large models (GPT-4o $60.0\%$ vs.\ $40.6\%$, Claude~3.5 Haiku $54.1\%$ vs.\ $27.2\%$, Gemini~2.0 Flash $59.5\%$ vs.\ $30.9\%$), for the reasoning model (GPT-5-mini $30.2\%$ vs.\ $14.8\%$), and for the small model (Llama-3-8B $53.7\%$ vs.\ $32.3\%$), as they do for humans ($47.6\%$ vs.\ $18.2\%$). This ordering is preserved by both comparison models; unlike in the private-value settings, GPT-5-mini makes substantial errors in common-value auctions. Appendix~\ref{app:winners_curse} reports, for the reference model, how winner losses vary with the number of bidders.}

\section{\texorpdfstring{\rev{Application to Design: Interventions and Market Structure}}{Application to Design: Interventions and Market Structure}}\label{session:ebay}

\rev{The laboratory benchmarks of Section~\ref{sec:benchmark_experimental} establish where LLM bidders agree with human bidders: at the ordinal tier, for the non-reasoning large model family. This section asks whether that agreement is useful for design. Before running a human experiment, a designer typically wants to know two things: whether a proposed intervention is likely to help, and whether one market structure is likely to induce better behavior than another. We use the panel to answer both. First, we treat the ordinal result as a prediction: a presentation change that improves human play should improve play for the non-reasoning large models, and we test one such change, the clock framing of a sealed-bid second-price auction \citep{breitmoser2022obviousness}. Second, we move from stylized laboratory mechanisms to a market structure documented in field data, an eBay-style proxy-bidding marketplace, and ask whether the panel recovers the ordering of bidding dynamics across closing rules.}

\subsection{\texorpdfstring{\rev{Clock framing as an intervention: a prediction test}}{Clock framing as an intervention: a prediction test}}\label{session:Intervention}

\rev{Recent studies ask whether behavior in strategy-proof auctions depends on how the mechanism is \emph{presented}, holding the incentives fixed. \citet{breitmoser2022obviousness} show that reframing a sealed-bid second-price auction as a dynamic ascending clock substantially increases truth-telling among human subjects, consistent with the idea that step-by-step ``stay/exit'' choices reduce the cognitive burden of mapping a value into a single sealed bid.\footnote{\citet{gonczarowski2023strategyproofness} study a second presentation change, an extensive-form \emph{menu} description of the same auction, and find no reliable effect on human bidding. We do not use it here because a null human effect makes no ordinal prediction.} If the non-reasoning large models preserve the ordering of human difficulty, the clock framing should raise their truthful play as well; the reasoning model, already at the benchmark, offers no room for improvement. We implement the clock framing as a change of prompt only: the allocation and payment rule are unchanged (Section~\ref{sec:data_generation}; prompts in Appendix~\ref{app:intervention}). Across all conditions we summarize behavior by the per-bid scaled deviation $(b-v)/\mathbb{E}[b^\star]$, with the same scaling as SMAD; negative values are underbids and positive values overbids.\footnote{An earlier version of this section reported a per-auction statistic (the deviation of the second-highest bid from the second-highest value) on a separate GPT-4o run, and correspondingly larger effect sizes; the qualitative pattern is unchanged. The cells reported here are the GPT-4o cells of Table~\ref{tab:clock_framing_main}, generated with the same templates as the other models.}}

\rev{For the reference model the prediction holds. Under the traditional description, GPT-4o bids are systematically below value (mean deviation $-0.118$, s.d.\ $0.128$, $n=300$). Under the clock framing the mean deviation moves to $-0.062$ (a $47\%$ reduction in average underbidding), the dispersion falls from $0.128$ to $0.084$, and the share of bids within $\pm 2\%$ of value rises from $28\%$ to $45\%$; both a Welch $t$-test and a Mann--Whitney test reject equality at $p<10^{-5}$ (Cohen's $d=0.48$). Table~\ref{tab:clock_framing_main} extends the test to the other two non-reasoning large model families, run with the same templates. The clock framing lowers SMAD for Claude~3.5 Haiku ($8.1\to7.1\%$) as it does for GPT-4o ($11.8\to6.2\%$), and raises it for Gemini~2.0 Flash ($5.0\to9.3\%$). The exception is instructive: Gemini's baseline second-price play is already the closest to truthful of the three ($62\%$ of bids within $\pm 2\%$ of value), so it is the family nearest the ceiling in this format, where the ordinal argument predicts no gain. Appendix~\ref{app:clock_robustness} repeats the test for an open-weights model outside the main panel, Gemma-3-27B, whose SMAD falls from $26.3\%$ to $13.5\%$. Claude's case is also worth reading closely: its mean absolute deviation falls only slightly, while the share of exactly truthful bids drops from $28\%$ to $7\%$. The clock framing removes Claude's overbidding in the human direction and replaces it with mild, uniform underbidding, a change in direction more than in magnitude. The prediction that a lever which helps human bidders helps the non-reasoning large models is thus borne out for the models with non-degenerate baseline error and fails, as it should, where there is no error to remove.}


\begin{table}[htbp]
\centering
\small
\rev{\resizebox{\linewidth}{!}{\begin{tabular}{l rr rr c l c}
\toprule
 & \multicolumn{2}{c}{Baseline} & \multicolumn{2}{c}{Clock framing} & within $\pm2\%$ & & matches \\
\cmidrule(lr){2-3}\cmidrule(lr){4-5}
Model & SMAD & mean dev & SMAD & mean dev & baseline $\to$ clock & $\Delta$ (MW) & human sign? \\
\midrule
GPT-4o & 11.8\% & $-0.118$ & 6.2\% & $-0.062$ & 28\% $\to$ 45\% & $\downarrow$ {\scriptsize $p<0.001$} & $\surd$ \\
Claude 3.5 Haiku & 8.1\% & $+0.022$ & 7.1\% & $-0.042$ & 28\% $\to$ 7\% & $\downarrow$ {\scriptsize $p<0.001$} & $\surd$ \\
Gemini 2.0 Flash & 5.0\% & $-0.050$ & 9.3\% & $-0.093$ & 62\% $\to$ 37\% & $\uparrow$ {\scriptsize $p<0.001$} & $\times$ \\
\bottomrule
\end{tabular}
}}
\caption{\rev{\textbf{Clock framing of the second-price sealed-bid auction across the three non-reasoning large model families.} SMAD (\%), mean scaled deviation $(b-v)/\mathbb{E}[b^\star]$, and the share of bids within $\pm2\%$ of value under the baseline description (two pooled runs of the standard description) and under the clock framing; $p$-values are Mann--Whitney tests against the same model's baseline. The final column records whether the sign of the effect matches the human evidence of \citet{breitmoser2022obviousness}: $\surd$ if SMAD falls with $p<0.05$, $\times$ otherwise.}}
\label{tab:clock_framing_main}
\end{table}

\subsection{\texorpdfstring{\rev{Market structure: closing rules in an eBay-style marketplace}}{Market structure: closing rules in an eBay-style marketplace}}\label{sec:ebay_market}

Having tested LLM behavior against human behavior in controlled laboratory experiment settings, we study a more complicated auction setting modeled after eBay's  auction-based marketplace. eBay's auction platform, studied by \citet{einav2011learning} and others, features design elements, \rev{for example proxy bidding and a fixed closing time}, that 
have been shown empirically to have substantial effects on bidder behavior. 
These differences highlight the gap between \textit{lab experiment} and \textit{real-world bidding}, where seemingly small design details can induce large shifts in timing, competition, and price discovery.

We implement a simplified eBay-style environment that isolates \rev{one design parameter documented in the empirical literature: the auction \emph{closing rule}}. The core mechanics of an eBay auction are simple: bidders on eBay use a proxy bidding system, which means that each bidder submits their ``maximum bid" to the platform, and the platform automatically bids on their behalf, only as much as necessary to keep the winning bidder in the lead (i.e., the second-highest bid + \$1, with incremental bidding managed automatically).
Bidders can increase their maximum bid at any point in time, until some closing time of 
the auction, giving this the flavor
of an open, second-price auction.
\rev{As described in Appendix~\ref{app:ebay_env}, we compare two treatments: (T1) the standard eBay hard close with proxy bidding, and (T2) proxy bidding with a soft-close (auction-extension) rule in the spirit of early Amazon auctions. The question is not whether LLM bidders reproduce the timing of human bids, but whether they recover the \emph{ordering} of bidding dynamics across the two closing rules that the field evidence documents: more last-moment bidding under a hard close than under a soft close.}

\paragraph{\rev{LLMs exhibit bid sniping under hard-close rules, and respond to soft-close extensions.}}\label{sec:ebay_sniping}

A hallmark of hard-close online auctions is \emph{bid sniping}: bidders delay their bids until the final moments of the auction to avoid triggering competitive responses . Sniping is widely observed on eBay \footnote{For example: \url{https://www.esnipe.com/}.}, and has long been studied as a behavioral and strategic departure from the simple ``bid-your-value'' logic implied by proxy bidding \citep{roth2002last}.
From the standpoint of market design, sniping is potentially costly because it compresses information revelation into the final moments of the auction, weakening timely price discovery. 
A classic intervention is a \emph{soft-close} (auction-extension) rule: if a new maximum bid arrives near the end, the auction is extended, allowing additional responses. Empirically, this design detail is credited with substantial differences in bidding dynamics between early Amazon.com auctions (soft close) and eBay (hard close).
Specifically, \citet{roth2002last} and \citet{ockenfels2006late} show that much less sniping occurred on Amazon.com, with bids on Amazon.com occurring throughout the auction compared to the empirical evidence of considerable bid
sniping on eBay.

\begin{figure}[tbp]
    \centering
    \includegraphics[width=0.85\linewidth]{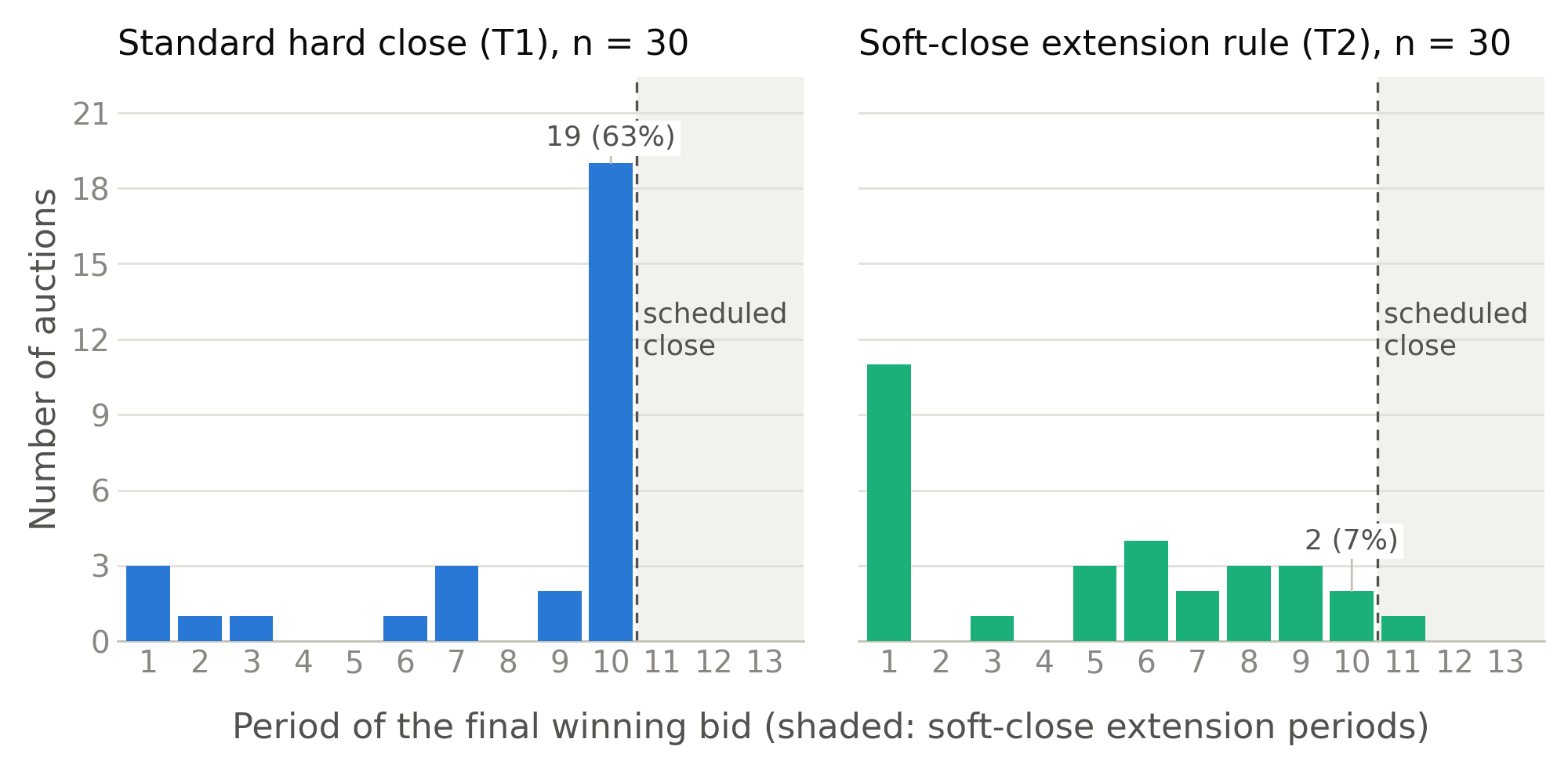}
    \caption{\rev{\textbf{Final winning-bid timing in simulated eBay-style auctions, IPV environment: standard hard close (left) vs.\ soft-close extension rule (right).} The final winning bid is the last period in which the eventual winner raises her maximum bid; the dashed line marks the scheduled close. The soft-close rule moves the timing of the final winning bid substantially earlier.}}
    \label{fig:final_winning_bid_shift}
\end{figure}

We ask whether LLM bidders respond to these temporal incentives in the same direction. We measure the \emph{final-winning-bid time}, defined as the last period in which the eventual winner changes her maximum bid. Intuitively, this statistic captures when the winner is effectively determined. Figure~\ref{fig:final_winning_bid_shift} reports histograms of this timing under hard close versus soft close, \rev{and Appendix~\ref{app:ebay_extra} reports the corresponding cumulative distributions}.

Two patterns emerge. First, under the standard hard-close rule (T1), LLM bidders concentrate their decisive bid updates near the end of the auction, consistent with sniping behavior\rev{: in $63\%$ of hard-close auctions ($19$ of $30$) the final winning bid is placed in the last scheduled period, and the median final-winning-bid period is the last one}. Second, introducing a soft-close extension rule (T2) sharply attenuates this last-period concentration: the distribution of final-winning-bid times shifts earlier, and last-period decisive bids become substantially less frequent\rev{: $7\%$ of soft-close auctions ($2$ of $30$) are decided in the last scheduled period, one more in an extension period, and the median final-winning-bid period falls to $5.5$ (Mann--Whitney $p=0.0002$; Kolmogorov--Smirnov $p=0.0003$)}. In short, LLM bidders respond strongly to the closing rule in the same qualitative direction emphasized in the human empirical literature: soft close de-incentivizes waiting and spreads competition earlier in time, improving the timing of information revelation. \rev{The magnitude of the shift is not the object of interest, since there is no human anchor for the timing distribution in our stylized environment. The ordering of the two closing rules, however, matches the field evidence, which is what the ordinal reading of Section~\ref{sec:benchmark_experimental} predicts for a non-stylized setting.}

%

%

\section{Discussion}\label{sec:discussion}

\subsection{\texorpdfstring{\rev{What LLM simulation can and cannot deliver for mechanism design}}{What LLM simulation can and cannot deliver for mechanism design}}

\rev{Simulations are proliferating in high-stakes engineering disciplines, where iterative design and validation are performed long before prototypes are built. Traditional mechanism design research, by contrast, relies on expensive human-subject experiments or on confounded observational data. Our results locate the role that LLM simulation can play in between. In magnitude, LLM behavior is not human behavior, and in direction it is often the opposite of human behavior: the same second-price auction that induces overbidding in the laboratory induces underbidding in most models. Any use of LLM bidders that reads off magnitudes (predicted revenue, predicted efficiency loss, predicted overbidding rates) is not supported by our evidence. At the ordinal level, however, the picture is different. For the non-reasoning large models, whose baseline deviations are neither negligible nor overwhelming, the cross-mechanism ordering of difficulty, the sign of the clock advantage, the sign of the first- versus second-price common-value comparison, the sign of the clock-framing effect, and the ordering of last-minute bidding across closing rules all match the human evidence. These are exactly the questions a designer asks when choosing among candidate mechanisms or presentations, and they are questions that dominant-strategy theory alone cannot answer, since it is silent on how the same incentives presented differently will be played.}

\rev{We therefore propose that LLM bidders be used as an \emph{ordinal screening} instrument within a two-tier workflow. In the first tier, a panel of models is run on the candidate designs, and the designs are ranked, or the sign of a proposed change is estimated, at a cost of a few dollars per cell. In the second tier, the designs that survive screening are taken to a human experiment, whose purpose is to establish magnitudes and directions, which the panel cannot supply. Three practices follow from our results. Models should be checked against known human orderings before they are used to screen new designs: neither nearly exact equilibrium play nor large errors alone establish suitability. Each model's use of extended reasoning and observed baseline errors should be reported with its results. And agreement across the panel, rather than the behavior of any single model, should be the evidence that an ordering is stable.}

\rev{LLM agents also operate on a text-based reasoning trace that can be queried. With further validation, these traces may offer information about the cognitive constraints that underlie deviations from classical auction theory; in this paper we use them only as a diagnostic and rely on them for no claim.}
%
%
%
%
%
%
%

\subsection{\texorpdfstring{\rev{Practical advantages and the scope for mechanism design}}{Practical advantages and the scope for mechanism design}}

\rev{LLM experiments avoid participant recruitment and can be repeated while varying a prompt or an auction rule. This complements online human laboratories, which have already lowered the cost of controlled behavioral research \citep{horton2011online, oprea2024decisions, enke2024behavioral}. Controlled simulations can identify how design changes affect the simulated agents under the tested protocol. Whether those effects carry over to humans requires separate validation. The low API cost of the reported experiments (under \$400 for the GPT-4o experiments, against roughly \$15,000 for a comparable human study; \citealp{li2017obviously}) is therefore a practical advantage for exploration, not evidence that they replace human experiments.}

\subsection{\texorpdfstring{\rev{Limitations}}{Limitations}}

\rev{Two limitations bound the interpretation. First, the human benchmarks are reconstructions of published moments, not micro data. The bands we report measure how far an anchor can move within what the source statistics leave undetermined; they are not sampling bands on human behavior, and we accordingly use the reconstructions only for orderings, directions, and comparative statics. Second, the variation across simulated bidders is confined to their induced values: every bidder receives the same objective and the same motivation. This mirrors the induced-value design of the human experiments we benchmark against, but it leaves open how a panel would behave if bidders also varied along other dimensions, such as their attitude to risk. Adding such variation in ways that preserve the incentive structure of the human experiments is a natural extension.}
%
%

\section{Conclusion}\label{sec:conclusion}

\rev{This paper reports more than 1,000 auction experiments with LLM bidders across five models, benchmarked against human laboratory evidence that we made comparable across mechanisms by reconstructing published moments. LLM bidders do not replicate human auction behavior: their bid distributions differ in magnitude, and in the second-price auction most models err in the opposite direction from human subjects. The surprising finding is that, without task-specific fine-tuning or calibration to human bids, the three non-reasoning large models robustly preserve key orderings of auction difficulty. Clock implementations reduce deviations wherever data are adequate, and clock framing helps two of the three models. More accurate equilibrium play does not guarantee greater similarity to humans: the reasoning model's near-equilibrium private-value bids leave little variation in errors to compare, while the small model's large errors yield an inverted ranking. Both nevertheless preserve the first- versus second-price common-value ordering.}

\rev{Auctions thus illustrate a broader research agenda for LLM simulations of human behavior: determine which features of behavior a model preserves, distinguishing magnitudes, directions, and orderings. Our results establish an out-of-the-box baseline for this agenda, complementary to models specifically trained on human choices. Where orderings have been validated, an LLM panel may help screen candidate interventions before human experiments establish their effects. Whether this approach works in voting, matching, contract design, or other behavioral settings remains an empirical question. The contribution is a way to test the usefulness of a simulation for a particular comparison, rather than assume that a model is a general substitute for human participants.}

\paragraph{Disclosure.} Claude (Anthropic, Opus 4.6) was used to assist with plotting code and manuscript formatting to style guide requirements

%
%
%
%
%

\bibliographystyle{llm_auction}
\bibliography{llm_auction}

@article{binz2025foundation,
  title   = {A foundation model to predict and capture human cognition},
  author  = {Binz, Marcel and Akata, Elif and Bethge, Matthias and others},
  journal = {Nature},
  volume  = {644},
  number  = {8078},
  pages   = {1002--1009},
  year    = {2025},
  doi     = {10.1038/s41586-025-09215-4},
  url     = {https://www.nature.com/articles/s41586-025-09215-4}
}

@article{wei2022chain,
  title={Chain-of-thought prompting elicits reasoning in large language models},
  author={Wei, Jason and Wang, Xuezhi and Schuurmans, Dale and Bosma, Maarten and Xia, Fei and Chi, Ed and Le, Quoc V and Zhou, Denny and others},
  journal={Advances in neural information processing systems},
  volume={35},
  pages={24824--24837},
  year={2022}
}

@article{dutting2024optimal,
  title={Optimal auctions through deep learning: Advances in differentiable economics},
  author={D{\"u}tting, Paul and Feng, Zhe and Narasimhan, Harikrishna and Parkes, David C and Ravindranath, Sai Srivatsa},
  journal={Journal of the ACM},
  volume={71},
  number={1},
  pages={1--53},
  year={2024},
  publisher={ACM New York, NY}
}

@article{ravindranath2023data,
  title={Data market design through deep learning},
  author={Ravindranath, Sai Srivatsa and Jiang, Yanchen and Parkes, David C},
  journal={Advances in Neural Information Processing Systems},
  volume={36},
  pages={6662--6689},
  year={2023}
}

@inproceedings{feng2018deep,
  title={Deep learning for revenue-optimal auctions with budgets},
  author={Feng, Zhe and Narasimhan, Harikrishna and Parkes, David C},
  booktitle={Proceedings of the 17th international conference on autonomous agents and multiagent systems},
  pages={354--362},
  year={2018}
}

@article{wang2025bundleflow,
  title={BundleFlow: Deep Menus for Combinatorial Auctions by Diffusion-Based Optimization},
  author={Wang, Tonghan and Jiang, Yanchen and Parkes, David C},
  journal={arXiv preprint arXiv:2502.15283},
  year={2025}
}

@article{fish2024algorithmic,
  title={Algorithmic Collusion by Large Language Models},
  author={Fish, Sara and Gonczarowski, Yannai A and Shorrer, Ran I},
  journal={arXiv preprint arXiv:2404.00806},
  year={2024}
}

@inproceedings{duetting2024mechanism,
  title={Mechanism design for large language models},
  author={Duetting, Paul and Mirrokni, Vahab and Paes Leme, Renato and Xu, Haifeng and Zuo, Song},
  booktitle={Proceedings of the ACM Web Conference 2024},
  pages={144--155},
  year={2024}
}

@article{jiang2025incentive,
  title={Incentive-Aligned Multi-Source LLM Summaries},
  author={Jiang, Yanchen and Feng, Zhe and Mehta, Aranyak},
  journal={arXiv preprint arXiv:2509.25184},
  year={2025}
}

@article{wang2026llm,
  title={LLM Active Alignment: A Nash Equilibrium Perspective},
  author={Wang, Tonghan and Pan, Yuqi and Yang, Xinyi and Jiang, Yanchen and Tambe, Milind and Parkes, David C},
  journal={arXiv preprint arXiv:2602.06836},
  year={2026}
}

@inproceedings{wang2024gemnet,
  title={GemNet: Menu-Based, Strategy-Proof Multi-Bidder Auctions Through Deep Learning},
  author={Wang, Tonghan and Jiang, Yanchen and Parkes, David C},
  booktitle={Proceedings of the 25th ACM Conference on Economics and Computation},
  pages={1100--1100},
  year={2024}
}

@article{soumalias2025llm,
  title={LLM-Powered Preference Elicitation in Combinatorial Assignment},
  author={Soumalias, Ermis and Jiang, Yanchen and Zhu, Kehang and Curry, Michael and Seuken, Sven and Parkes, David C},
  journal={arXiv preprint arXiv:2502.10308},
  year={2025}
}

@article{curry102automated,
  title={Automated mechanism design: A survey},
  author={Curry, Michael J. and Fan, Zhou and Jiang, Yanchen and Ravindranath, Sai Srivatsa and Wang, Tonghan and Parkes, David C.},
  journal      = {ACM SIGecom Exchanges},
  volume       = {22},
  number       = {2},
  year         = {2025},
  pages        = {102-120},    
}

@article{huang2025accelerated,
  title={Accelerated Preference Elicitation with LLM-Based Proxies},
  author={Huang, David and Marmolejo-Coss{\'\i}o, Francisco and Lock, Edwin and Parkes, David},
  journal={arXiv preprint arXiv:2501.14625},
  year={2025}
}

@article{kagel1993independent,
  title={Independent private value auctions: Bidder behaviour in first-, second-and third-price auctions with varying numbers of bidders},
  author={Kagel, John H and Levin, Dan},
  journal={The Economic Journal},
  volume={103},
  number={419},
  pages={868--879},
  year={1993},
  publisher={Oxford University Press Oxford, UK}
}

@article{horton2011online,
  title={The online laboratory: Conducting experiments in a real labor market},
  author={Horton, John J and Rand, David G and Zeckhauser, Richard J},
  journal={Experimental economics},
  volume={14},
  pages={399--425},
  year={2011},
  publisher={Springer}
}

@article{oprea2024decisions,
  title={Decisions under risk are decisions under complexity},
  author={Oprea, Ryan},
  journal={American Economic Review},
  volume={114},
  number={12},
  pages={3789--3811},
  year={2024},
  publisher={American Economic Association 2014 Broadway, Suite 305, Nashville, TN 37203}
}

@article{roth2002last,
  title={Last-minute bidding and the rules for ending second-price auctions: Evidence from eBay and Amazon auctions on the Internet},
  author={Roth, Alvin E and Ockenfels, Axel},
  journal={American economic review},
  volume={92},
  number={4},
  pages={1093--1103},
  year={2002},
  publisher={American Economic Association}
}

@techreport{enke2024behavioral,
  title={Behavioral attenuation},
  author={Enke, Benjamin and Graeber, Thomas and Oprea, Ryan and Yang, Jeffrey},
  year={2024},
  institution={National Bureau of Economic Research}
}

@techreport{einav2011learning,
  title={Learning from seller experiments in online markets},
  author={Einav, Liran and Kuchler, Theresa and Levin, Jonathan D and Sundaresan, Neel},
  year={2011},
  institution={National Bureau of Economic Research}
}

@book{kagel2020handbook,
  title={The handbook of experimental economics, volume 2},
  author={Kagel, John H and Roth, Alvin E},
  year={2020},
  publisher={Princeton university press}
}

@article{coppinger1980incentives,
  title={incentives and behavior in english, dutch and sealed-bid auctions},
  author={Coppinger, Vicki M and Smith, Vernon L and Titus, Jon A},
  journal={Economic inquiry},
  volume={18},
  number={1},
  pages={1--22},
  year={1980},
  publisher={Wiley Online Library}
}

@article{cox1988theory,
  title={Theory and individual behavior of first-price auctions},
  author={Cox, James C and Smith, Vernon L and Walker, James M},
  journal={Journal of Risk and uncertainty},
  volume={1},
  pages={61--99},
  year={1988},
  publisher={Springer}
}

@article{kagel1987information,
  title={Information impact and allocation rules in auctions with affiliated private values: A laboratory study},
  author={Kagel, John H and Harstad, Ronald M and Levin, Dan and others},
  journal={Econometrica},
  volume={55},
  number={6},
  pages={1275--1304},
  year={1987}
}

@article{myerson1981optimal,
  title={Optimal auction design},
  author={Myerson, Roger B},
  journal={Mathematics of operations research},
  volume={6},
  number={1},
  pages={58--73},
  year={1981},
  publisher={INFORMS}
}

@article{bazerman1983won,
  title={I won the auction but don't want the prize},
  author={Bazerman, Max H and Samuelson, William F},
  journal={Journal of conflict resolution},
  volume={27},
  number={4},
  pages={618--634},
  year={1983},
  publisher={Sage Publications 275 South Beverly Drive, Beverly Hills, California 90212}
}

@article{kagel1986winner,
  title={The winner's curse and public information in common value auctions},
  author={Kagel, John H and Levin, Dan},
  journal={The American economic review},
  pages={894--920},
  year={1986},
  publisher={JSTOR}
}

@article{li2017obviously,
  title={Obviously strategy-proof mechanisms},
  author={Li, Shengwu},
  journal={American Economic Review},
  volume={107},
  number={11},
  pages={3257--3287},
  year={2017},
}

@article{breitmoser2022obviousness,
  title={Obviousness around the clock},
  author={Breitmoser, Yves and Schweighofer-Kodritsch, Sebastian},
  journal={Experimental Economics},
  volume={25},
  number={2},
  pages={483--513},
  year={2022},
  publisher={Springer}
}

@article{bubeck2023sparks,
  title={Sparks of artificial general intelligence: Early experiments with gpt-4},
  author={Bubeck, S{\'e}bastien and Chandrasekaran, Varun and Eldan, Ronen and Gehrke, Johannes and Horvitz, Eric and Kamar, Ece and Lee, Peter and Lee, Yin Tat and Li, Yuanzhi and Lundberg, Scott and others},
  journal={arXiv preprint arXiv:2303.12712},
  year={2023}
}

@techreport{horton2023large,
  title={Large language models as simulated economic agents: What can we learn from homo silicus?},
  author={Horton, John J},
  year={2023},
  institution={National Bureau of Economic Research}
}

@techreport{manning2024automated,
  title={Automated social science: Language models as scientist and subjects},
  author={Manning, Benjamin S and Zhu, Kehang and Horton, John J},
  year={2024},
  institution={National Bureau of Economic Research}
}

@techreport{Horton2024EDSL,
title = {EDSL: Expected Parrot Domain Specific Language for AI Powered Social Science},
author = {John Horton and Apostolos Filippas and Robin Horton},
year = {2024},
institution = {Expected Parrot},
type = {Whitepaper}
}

@inproceedings{aher2023using,
  title={Using large language models to simulate multiple humans and replicate human subject studies},
  author={Aher, Gati V and Arriaga, Rosa I and Kalai, Adam Tauman},
  booktitle={International Conference on Machine Learning},
  pages={337--371},
  year={2023},
  organization={PMLR}
}

@article{achiam2023gpt,
  title={Gpt-4 technical report},
  author={Achiam, Josh and Adler, Steven and Agarwal, Sandhini and Ahmad, Lama and Akkaya, Ilge and Aleman, Florencia Leoni and Almeida, Diogo and Altenschmidt, Janko and Altman, Sam and Anadkat, Shyamal and others},
  journal={arXiv preprint arXiv:2303.08774},
  year={2023}
}

@article{wan2023kelly,
  title={" kelly is a warm person, joseph is a role model": Gender biases in llm-generated reference letters},
  author={Wan, Yixin and Pu, George and Sun, Jiao and Garimella, Aparna and Chang, Kai-Wei and Peng, Nanyun},
  journal={arXiv preprint arXiv:2310.09219},
  year={2023}
}

@article{chen2023put,
  title={Put your money where your mouth is: Evaluating strategic planning and execution of llm agents in an auction arena},
  author={Chen, Jiangjie and Yuan, Siyu and Ye, Rong and Majumder, Bodhisattwa Prasad and Richardson, Kyle},
  journal={arXiv preprint arXiv:2310.05746},
  year={2023}
}

@article{daskalakis2024charting,
  title={Charting the Shapes of Stories with Game Theory},
  author={Daskalakis, Constantinos and Gemp, Ian and Jiang, Yanchen and Leme, Renato Paes and Papadimitriou, Christos and Piliouras, Georgios},
  journal={arXiv preprint arXiv:2412.05747},
  year={2024}
}

@inproceedings{park2023generative,
  title={Generative agents: Interactive simulacra of human behavior},
  author={Park, Joon Sung and O'Brien, Joseph and Cai, Carrie Jun and Morris, Meredith Ringel and Liang, Percy and Bernstein, Michael S},
  booktitle={Proceedings of the 36th Annual ACM Symposium on User Interface Software and Technology},
  pages={1--22},
  year={2023}
}

@book{krishna2009auction,
  title={Auction theory},
  author={Krishna, Vijay},
  year={2009},
  publisher={Academic press}
}

@article{ockenfels2006late,
  title={Late and multiple bidding in second price Internet auctions: Theory and evidence concerning different rules for ending an auction},
  author={Ockenfels, Axel and Roth, Alvin E},
  journal={Games and Economic behavior},
  volume={55},
  number={2},
  pages={297--320},
  year={2006},
  publisher={Elsevier}
}

@techreport{gonczarowski2023strategyproofness,
  title={Strategyproofness-exposing mechanism descriptions},
  author={Gonczarowski, Yannai A and Heffetz, Ori and Thomas, Clayton},
  year={2023},
  institution={National Bureau of Economic Research}
}

@article{levin1996revenue,
  title={Revenue effects and information processing in English common value auctions},
  author={Levin, Dan and Kagel, John H and Richard, Jean-Francois},
  journal={The American Economic Review},
  pages={442--460},
  year={1996},
  publisher={JSTOR}
}

@inproceedings{dong2024survey,
  title={A survey on in-context learning},
  author={Dong, Qingxiu and Li, Lei and Dai, Damai and Zheng, Ce and Ma, Jingyuan and Li, Rui and Xia, Heming and Xu, Jingjing and Wu, Zhiyong and Chang, Baobao and others},
  booktitle={Proceedings of the 2024 conference on empirical methods in natural language processing},
  pages={1107--1128},
  year={2024}
}

@article{infobid2025,
  title={InfoBid: A Simulation Framework for Studying Information Disclosure in Auctions with Large Language Model-based Agents},
  author={Zhang, Yue and others},
  journal={arXiv preprint arXiv:2503.22726},
  year={2025}
}

@article{lu2025claude,
  title={Claude, Gemini, and ChatGPT Walk into an Auction},
  author={Lu, Janna},
  journal={SSRN working paper},
  year={2025},
  note={Available at SSRN: https://papers.ssrn.com/sol3/papers.cfm?abstract\_id=5335262}
}

@article{tsuchihashi2023homo,
  title={Do AIs Dream of Homo Economicus? Answers from ChatGPT},
  author={Tsuchihashi, Toshihiro},
  journal={SSRN working paper},
  year={2023},
  note={Available at SSRN: https://papers.ssrn.com/sol3/papers.cfm?abstract\_id=4495500}
}

@article{corrigan2025usefulness,
  title={On the Usefulness of Using Current LLMs for Experimental Auction Valuation},
  author={Corrigan, Jay R and Grebitus, Carola and Rousu, Matthew C},
  journal={SSRN working paper},
  year={2025},
  note={Available at SSRN: https://papers.ssrn.com/sol3/papers.cfm?abstract\_id=5217193}
}

@article{auctionnet2024,
  title={AuctionNet: A Novel Benchmark for Decision-Making in Large-Scale Games},
  author={Chen, Zhiyu and others},
  journal={arXiv preprint arXiv:2412.10798},
  year={2024}
}

@inproceedings{ramage2024adauctions,
  title={Ad Auctions for LLMs via Retrieval Augmented Generation},
  author={MohammadTaghi Hajiaghayi and S{\'e}bastien Lahaie and Keivan Rezaei and Suho Shin},
  booktitle={Advances in Neural Information Processing Systems},
  volume={37},
  year={2024}
}

@article{smith1976experimental,
  author  = {Smith, Vernon L.},
  title   = {Experimental Economics: Induced Value Theory},
  journal = {American Economic Review},
  volume  = {66},
  number  = {2},
  pages   = {274--279},
  year    = {1976}
}

@article{smith1982microeconomic,
  author  = {Smith, Vernon L.},
  title   = {Microeconomic Systems as an Experimental Science},
  journal = {American Economic Review},
  volume  = {72},
  number  = {5},
  pages   = {923--955},
  year    = {1982}
}
\newpage
\appendix
\section{Details about Human Data Collection}\label{app:human}

\paragraph{Mapping original reported statistics into the unified \rev{deviation} metric.}

\textbf{(Li; Breitmoser--Schweighofer-Kodritsch) AC/AC-B/2P under APV.}
These papers report Mean Absolute Deviation (MAD)\footnote{For clock auctions, since the winner did not submit a final sealed bid, the reported MAD is computed only from bidders who revealed a bid by dropping out (i.e., those who exited the clock), excluding the winner's unobserved terminal willingness-to-pay.}
 from truthful bidding. In Li (2017), the reported deviation is based on a group-level statistic involving the second-highest bid and second-highest value; in Breitmoser--Schweighofer-Kodritsch, deviations are computed from submitted bids (often focusing on non-winning bids). In each case we adhere to the paper's unit of observation, but the SMAD definition conceptually applies to the underlying submitted bids.
Because the equilibrium optimum under AC and 2P is truthful bidding $b^\star(v)=v$, we have $d_m=|b-v|$ and thus
\[
\Delta_m = 100\cdot \frac{\mathrm{MAD}_m}{\mu_m^\star},
\qquad
\mu_m^\star=\mathbb E[v].
\]
Here $\mathbb E[v]$ is implied by the experimental value distribution reported by the paper (e.g., $\mathbb E[v]=70$ in Li; and $\mathbb E[v]=65$ in Breitmoser--Schweighofer-Kodritsch given their stated common-value and idiosyncratic components).\footnote{When auctions are repeated, we follow the paper's convention and report post-learning blocks (e.g., dropping the first three rounds and using the next reported block).}
If the paper reports $\mathrm{SE}(\mathrm{MAD}_m)$ (often using group means as the unit of observation), we propagate it via
\[
\mathrm{SE}(\widehat{\Delta}_m)=100\cdot \frac{\mathrm{SE}(\widehat{\mathrm{MAD}}_m)}{\widehat{\mu}_m^\star},
\quad
\mathrm{CI}_{95\%}(\Delta_m)=\widehat{\Delta}_m \pm 1.96\,\mathrm{SE}(\widehat{\Delta}_m).
\]

\textbf{(Kagel--Levin 1993) First-/Second-/Third-price IPV.}
Kagel--Levin report bid regressions and over-/under-bidding frequencies for fixed bidder counts (e.g., $N\in\{5,10\}$). We treat each $N$ separately: the regression form $\hat b(x)=\hat\alpha+\hat\beta x$ is fit for a \emph{fixed} number of bidders, and is not pooled across varying $N$.
Theory gives linear equilibrium bid functions under IPV with $x\sim U[x_{\min},x_{\max}]$:

$$\begin{aligned}
b^\star_{\mathrm{FP\_IPV}}(x) &= x - \frac{x-x_{\min}}{N} = \frac{N-1}{N}x + \frac{1}{N}x_{\min} \\
b^\star_{\mathrm{SP\_IPV}}(x) &= x \\
b^\star_{\mathrm{TP\_IPV}}(x) &= x + \frac{x-x_{\min}}{N-2} = \frac{N-1}{N-2}x - \frac{1}{N-2}x_{\min}
\end{aligned}$$

(When $x_{\min}=0$, these reduce to the familiar slope-only formulas.)
Because only summary moments are reported, we reconstruct the bid distribution via a parametric data-generating process (DGP) consistent with Table~2.
For each auction series and fixed $N$, we assume $x\sim U[x_{\min},x_{\max}]$ (bounds from the paper), and generate bids as
\[
b = \hat\alpha+\hat\beta x + \varepsilon,\qquad \varepsilon\sim \mathcal N(\mu,\sigma^2),
\]
where $(\mu,\sigma)$ are fitted by moment matching to the reported frequencies (e.g., $\Pr(b>x)$ and $\Pr(b>b^\star(x))$).
Given the fitted DGP, we estimate
\[
\widehat{\mathrm{MAD}}_m = \mathbb E\!\left[\,|b-b^\star(x)|\,\right],
\qquad
\widehat{\mu}_m^\star=\mathbb E[b^\star(x)],
\qquad
\widehat{\Delta}_m = 100\cdot \frac{\widehat{\mathrm{MAD}}_m}{\widehat{\mu}_m^\star}.
\]
In practice, we run the Monte Carlo procedure 100 times and report the across-run mean $\widehat{\Delta}_m$ as the point estimate; uncertainty is obtained from across-run variability (percentile or normal approximation).

\textbf{(Levin--Kagel--Richard 1996 and related CV papers) First-price CV and English (second-price proxy).}
In common-value auctions, bidders observe signals $s_i$ about a common value $V$ (e.g., $s_i=V+\varepsilon_i$ as specified in the original paper). 

Theory gives linear equilibrium bid functions under Common Value:
$$\begin{aligned}
b^*_{FP\_CV}(x) &= x - \epsilon + Y\\
Y &= \left( \frac{2\epsilon}{N+1} \right) \exp\left[ -\left( \frac{N}{2\epsilon} \right) (x - (\underline{x} + \epsilon)) \right]\\
b^*_{SP\_CV}(x) &= x
\end{aligned}$$

The theoretical benchmark is the equilibrium bidding strategy $b_m^\star(s)$ implied by the paper's signal structure and mechanism. We therefore compute SMAD \emph{on bids} exactly as in private values:
\[
d_m = |b - b_m^\star(s)|,
\qquad
\Delta_m = 100\cdot \frac{\mathbb E[|b-b_m^\star(s)|]}{\widehat{\mu}_m^\star}.
\]

Because there is a very large common part in each players' valuation, we take the following normalization factor:
\[
{\widehat{\mu}_m^\star} = \mathbb E[b^\star_{max}(x) - b^\star_{min}(x)]
\]

When micro-data are unavailable, we reconstruct the bid distribution by simulating the paper's DGP (signals and value draws) together with the theoretical bidding rule $b_m^\star(\cdot)$, and fit any auxiliary noise parameters by moment matching to the summary statistics reported (e.g., frequencies of overbidding/high bids, or mean prices/profits as available). We run Monte Carlo 100 times and report the across-run mean $\widehat{\Delta}_m$ as the point estimate; uncertainty is obtained from across-run variability (percentile or normal approximation).

\subsection{Mixture Model Fitting Details}\label{app:mixture}

\begin{table}[h]
\centering
\small
\begin{tabular}{llcccc}
\toprule
Source & Treatment & $p_{\text{eq}}$ & $p_{\text{over}}$ & $\lambda$ & Matched moment \\
\midrule
Li (2017) & 2P & 0.50 & 0.40 & 0.219 & Mean ratio = 1.08 \\
Li (2017) & AC & 0.67 & 0.18 & 0.198 & Mean ratio = 1.02 \\
Breitmoser (2022) & 2P & 0.44 & 0.40 & 0.283 & Mean ratio = 1.10 \\
Breitmoser (2022) & AC & 0.83 & 0.12 & 0.154 & Mean ratio = 1.01 \\
Kagel-Levin (1993) & FPSB, $n$=5 & 0.75 & 0.12 & 0.100 & $R^2$=0.88 \\
Kagel-Levin (1993) & SPSB, $n$=5 & 0.27 & 0.67 & 0.153 & $R^2$=0.97 \\
Kagel-Levin (1993) & TPSB, $n$=5 & 0.58 & 0.27 & 0.924 & $R^2$=0.94 \\
Gonczarowski (2022) & Traditional & 0.37 & 0.41 & 0.450 & MAD = \$0.51 \\
Gonczarowski (2022) & Menu & 0.34 & 0.43 & 0.450 & MAD = \$0.55 \\
\bottomrule
\end{tabular}
\caption{Fitted mixture model parameters by source and treatment. Mixture weights from reported rates; $\lambda$ chosen to match the secondary statistic listed; $p_{\text{under}} = 1 - p_{\text{eq}} -
p_{\text{over}}$.}
\label{tab:mixture_fit}
\end{table}

\subsection{\texorpdfstring{\rev{Reconstruction-Uncertainty Bands}}{Reconstruction-Uncertainty Bands}}\label{app:bands}

\rev{Beyond the Monte Carlo simulation noise at fixed mixture parameters, we represent uncertainty in the fitted parameters themselves by a parametric bootstrap over the mixture parameters. For each source--treatment pair we draw $B=2000$ vectors of $(p_{\text{eq}}, p_{\text{over}}, p_{\text{under}}, \lambda, \sigma, \delta)$ from admissible ranges pinned to the reported statistics and their granularity, propagate each draw through the reconstruction that defines the anchor, and report percentile bands. The point anchors are reproduced exactly at the central (published) parameter values before any perturbation, so the bands are centered on the numbers quoted in the main text. The admissible ranges are as follows. The direction weights $(p_{\text{eq}}, p_{\text{over}}, p_{\text{under}})$ are re-drawn from the sampling uncertainty of the reported over-/under-/at-value frequencies: a multinomial resample of the reported cell counts for \citet{kagel1993independent}, and a $\pm 5$ percentage-point range around each reported rate for \citet{li2017obviously} and \citet{breitmoser2022obviousness}, whose rates are stated only to that granularity (``${\sim}50\%$ dominant-strategy play'' admits $45$--$55\%$). The mean bid-to-value ratio is re-drawn through the reported bid-regression slope within its standard error ($\mathrm{SE}\approx 0.01$), which is the dominant driver of the sealed-bid IPV bands: for first-price the anchor is the bid-shading gap between human bids, which sit essentially at value, and the risk-neutral equilibrium $\tfrac{N-1}{N}v$. The equilibrium noise $\sigma$ is re-drawn through the reported $R^2$ within its reporting precision. The underbid bound $\delta$ is re-drawn over the range $[0.10, 0.25]$ used in the reconstruction. For the clock and APV formats the reported mean absolute deviation is re-drawn within its published standard error, which reproduces the confidence intervals carried in Table~\ref{tab:mixture_fit}. The overbid scale $\lambda$ governs the shape rather than the level of the deviation mass and is pinned so that the central mixture reproduces the published anchor.}

\rev{These are reconstruction bands, not sampling bands on human behavior: they quantify how far the reconstructed anchor can move within what the source statistics leave undetermined. Table~\ref{tab:bands} reports them. Under the bands, the two load-bearing human orderings are stable: the first-price $\gg$ second-price difficulty gap holds in $100\%$ of re-draws (SMAD ratio $95\%$ band $[3.54, 5.65]$, centered on the $4.4\times$ point gap), and the clock improvement over sealed second-price bidding holds in $100\%$ of re-draws for the open clock (AC) and in $92.6\%$ for the closed clock (AC-B), the latter's wider band inherited from the large reported standard error on the \citet{breitmoser2022obviousness} mean absolute deviation. Because the reconstructions carry this uncertainty, we use them only for orderings, directions, and comparative statics, never for level comparisons or hypothesis tests against LLM cells. The bootstrap is implemented in \texttt{analysis/reconstruction\_bands.py}.}

\begin{table}[htbp]
\centering
\rev{\resizebox{\linewidth}{!}{\begin{tabular}{llrcc}
\toprule
Format & Source & Point & 68\% band & 95\% band \\
\midrule
First-Price IPV & \citet{kagel1993independent} & 24.76 & [23.55, 25.97] & [22.30, 27.25] \\
Second-Price IPV & \citet{kagel1993independent} & 5.65 & [5.03, 6.23] & [4.46, 6.79] \\
Third-Price IPV ($n=5$) & \citet{kagel1993independent} & 7.66 & [6.92, 8.39] & [6.15, 9.15] \\
Second-Price APV & \citet{li2017obviously} & 9.31 & [7.99, 10.56] & [6.78, 11.74] \\
Ascending Clock APV & \citet{li2017obviously} & 3.54 & [2.69, 4.37] & [2.00, 5.36] \\
AC-Closed (AC-B) APV & \citet{breitmoser2022obviousness} & 5.83 & [3.77, 7.75] & [2.30, 10.11] \\
\bottomrule
\end{tabular}
}}
\caption{\rev{Reconstruction-uncertainty bands on the human SMAD anchors (parametric bootstrap over the fitted mixture parameters, $B=2000$; 68\% and 95\% percentile bands). Common-value anchors have no reconstruction band because they are profit-based moments taken directly from the source.}}
\label{tab:bands}
\end{table}

\section{LLM Simulation Procedure}

We give the prompt-and-parse pipeline used to simulate various auctions with LLM agents. Each simulated bidder is queried independently. In every auction instance, the agent (i) receives the scenario and mechanism description, (ii) is asked to articulate a short bidding plan, and (iii) outputs a bid/decision in a structured format that can be parsed deterministically.

\subsection{Simulation process in sealed-bid auctions}\label{app:process}

\paragraph{Prompt structure.}
We attach the canonical prompt template in the following. Each bidder is assigned a role (e.g., ``Bidder Andy'') and informed of the other bidders' identities. The prompt then provides (a) a mechanism-specific \emph{rule explanation} (e.g., first-price vs.\ second-price vs.\ third-price), and (b) universal \emph{instructions} that define the objective and response format. We require the model to respond using two exact tags: \verb|<PLAN>| (a brief natural-language rationale) and \verb|<ACTION>| (the bid as a number) in the Chain-of-Thought style \citep{wei2022chain}:

\begin{figure}[H]
\centering
\begin{minipage}{0.95\textwidth}
\begin{lstlisting}
    You are Bidder Andy.            
    You are bidding with {OTHER_BIDDERS}.     
    
    {RULE EXPLANATION} + {INSTRUCTIONS}

    Your response must use these EXACT tags below:
    ```
    <PLAN>[ Reason out a bidding strategy and plan for this round. {% if current_round > 1 %}Learn from your previous rounds. Remember: the values will be different in future rounds - don't assume patterns will repeat. {% else %}Consider the history above if available. {% endif %}Be detailed and precise. LIMIT your plan to 100 words. ] </PLAN>
    <ACTION> [Your bid amount as a number, e.g., 1, 44, 50] </ACTION>
    ```
\end{lstlisting}
\end{minipage}
\end{figure}

\paragraph{Universal objective instruction.}
Across all sealed-bid simulations, we append a universal instruction block that specifies the optimization objective (maximize profit). \rev{In the one-shot baseline the block reads: ``You have only one opportunity to bid. Your top priority is to place a bid that maximizes your expected profit.'' In the repeated-round learning ablation (Appendix~\ref{app:learning}) it is replaced by the multi-round text below, which additionally discourages spurious cross-round pattern extrapolation \citep{fish2024algorithmic}.}
\begin{figure}[H]
\centering
\begin{minipage}{0.95\textwidth}
\begin{lstlisting}
    Your TOP PRIORITY is to place bids which maximize the user's profit in the long run. 
    Learn from the history of previous rounds in order to maximize your total profit. 
    Don't forget the values are redrawn independently each round.
\end{lstlisting}
\end{minipage}
\end{figure}

\paragraph{History and feedback (multi-round ablation only).}
Our main experiments restrict attention to single-round auctions ($T=1$). When we run the multi-round ablation in Appendix~\ref{app:learning}, we include a standardized history transcript in the prompt. The history contains the agent's past realized values, past bids, whether it won, the transaction price, and realized profit, as well as the list of all bids in each prior round. This transcript is appended above the response-tag requirements, so that the agent can condition on past outcomes without ambiguity.

\subsection{Simulation process in Ascending Clock auctions}\label{app:dynamic}

In Ascending Clock Auctions, each round consists of multiple clock cycles, during which every bidder is asked whether they want to stay or drop out at the current clock price.
In the first round, bidders are reminded of the auction rules, similar to those outlined in Section \ref{app:process}. 
The auction starts with an initial price of 0, which increases incrementally until only one bidder remains or two bidders drop out simultaneously, in which case the winner is chosen randomly. 
The detailed prompt is listed as follows, with variables enclosed in brackets:

\begin{figure}[H]
\centering
\begin{minipage}{0.95\textwidth}
\begin{lstlisting}
    You are Bidder Andy. You are bidding with {OTHER_BIDDERS}.     
    
    {RULE EXPLANATION} + {INSTRUCTIONS} 

    Your value towards to the prize is 26 in this round.
    The current price in this clock cycle is {current_price}. 
    The price for next clock cycle is {current_price + increment}.

    The previous bidding history is: {transcript}.

    Do you want to stay in the bidding? 
    If you choose yes, you can keep bidding for next clock. If you choose No, you will exit and have no chance to re-enter the bidding. Your response must use these EXACT tags below. You must output the ACTION.
    ```
    <PLAN>
    [Write your plans for bidding strategies. Be detailed and precise but keep things succinct and don't repeat yourself. LIMIT your plan to 50 words. ] </PLAN>
    <ACTION> Yes or No </ACTION>
    ```
\end{lstlisting}
\end{minipage}
\label{fig:ac_prompt}
\end{figure}

In AC, if LLM bidders didn't decide to drop out in the first round, they are shown the previous bidding history for the next clock cycle. The history includes the price and how many people had dropped out. 
And the transcript variable will be the following format:
\textit{    The previous biddings are: 
    ['In clock round 1, the price was 1, 
    no players dropped out'].}

In AC-B, we don't show the previous biddings and in each clock cycle, we directly query LLM agents' decision. 
So the transcript variable will be None.

A clock auction round ends when only one bidder remains, who is declared the winner, or when two bidders drop out simultaneously, in which case the winner is chosen randomly. Each auction consists of 10 independent rounds with affiliated value settings.

\subsection{\rev{Stylized eBay proxy-bidding environment}}\label{app:ebay_env}
\rev{This section was part of the Methods in an earlier version; it is moved here because the eBay-style environment is separate from the laboratory benchmarks.}
Because our eBay-style results section studies a richer, marketplace-inspired microstructure rather than canonical lab mechanisms, we treat eBay simulations as a separate environment. We implement a stylized eBay auction with \emph{proxy bidding} and an optional auction-extension closing rule. Proxy bidding means each bidder maintains a \emph{maximum bid} (their willingness-to-pay cap), and the platform automatically raises the standing bid only as needed to keep the current leader ahead. In each eBay simulation, there are $n=3$ bidders and (unless modified by the closing rule) the auction lasts for $T_{\text{eBay}}=10$ discrete bidding periods. Unless otherwise noted, the eBay environment uses IPV with $v_i \sim \mathrm{Unif}[0,99]$.

\emph{Treatments (closing rule).}
\rev{We vary one design dimension, the closing rule, yielding two treatments: (T1) the standard eBay auction with proxy bidding and a hard close, and (T2) the same auction with a modified closing rule (auction extension) in the spirit of early Amazon auctions. An earlier version of this paper also varied a hidden reserve price; we omit it here to keep the environment minimal, and because it had no detectable effect on outcomes in our setting.}
%
%

\emph{Discrete-time implementation of bidding.}
Because LLM interactions are necessarily discrete, we approximate continuous-time bidding using period-based actions. In each period, bidders move in a common-knowledge, randomly permuted order. When it is a bidder's turn, they observe the current auction state (including updates induced by earlier movers in that period) and decide whether to \emph{increase their maximum bid} or \emph{hold}. In the final period, bidding is simultaneous: bidders submit their final decisions without observing others' last-period actions, capturing the idea that, in continuous time, a bidder cannot know whether another bid will arrive just before closing.

\emph{Modified closing rule (auction extension).}
Under the modified closing rule, if a higher maximum bid is placed in the final period such that a new bidder becomes the winner, the auction is extended by one additional period. This process repeats until a final period occurs in which no further bids are made.

\emph{Abstraction and prompting.}
This stylized eBay framework abstracts away from additional platform features such as shipping fees and Buy-It-Now options. Moreover, unlike the explicit framing interventions in Section~\ref{session:Intervention}, we keep the eBay prompts deliberately \emph{minimal} and descriptive: the goal is to study how LLMs behave under richer market microstructure without steering them toward any particular strategy through didactic instruction. The complete simulation process is described in Appendix~\ref{app:ebay}, and the corresponding prompts are listed in Appendix~\ref{app:prompt}.

\subsection{Simulation process in eBay auctions}\label{app:ebay}

For eBay auctions, we discretize the continuous bidding period into 10 periods. Each period, LLM agents decide whether to increase their bid or hold their current bid. The bidding process follows a structured format, where bidders act in a predefined sequence each period (e.g., Charles, then Alice, then Betty). Agents are informed of past price changes through a transcript, such as ``On day 1, the price changed to 1. On day 2, the price changed to 3.''

The prompt provides key details, including the bidder's private value, the total number of bidding days, the current day, the bidding order, previous bids, previous proxy bids, and the current price. The decision-making process is again guided by a structured asking prompt that enforces bidding rules. 
\begin{figure}[H]
\centering
\begin{minipage}{0.95\textwidth}
\begin{lstlisting}
    You are Bidder Andy.            
    You are bidding with {OTHER_BIDDERS}.     
    
    {RULE EXPLANATION} + {INSTRUCTIONS} 
    
    Your private value for this item is ${private_value}. This is the maximum amount you are willing to pay. Keep this value private. There are in total {total_periods} days  of bidding and this is day {current_period}. {ordering}.
    
    Your previous bids are {previous_bid}. If you have already placed bids, you can only increase your bid or hold your current bid. The previous proxy bids are: {transcript}.
    The current price is {current_price}.
    Your response must use these EXACT tags below. Don't miss the amount.
    ```
    <PLAN>[Write your plans for what bidding strategies to test next. Be detailed and precise but keep things succinct and don't repeat yourself. LIMIT your plan to 50 words. ]</PLAN>
    <CHECK>your last bid is {{last_bid_amount}}, you cannot bid lower that this value </CHECK>
    <ACTION> BID or HOLD  </ACTION>
    <AMOUNT> if BID, enter a number here, e.g. 1, 44. If HOLD, enter 0 </AMOUNT>
    ```
\end{lstlisting}
\end{minipage}
\label{fig:ebay_prompt}
\end{figure}

\section{Examining Learning Effects in Repeated Games}\label{app:learning}

To assess whether LLM agents exhibited learning or adaptation across rounds, we conduct a set of statisitcal tests comparing early-round behavior to late-round behavior and examining continuous trends over time. For the First-Price Sealed-Bid (FPSB) auction, the mean SMAD in the first five rounds (0.2523) is essentially identical to that in the last five rounds (0.2524), with no detectable difference in either a t-test ($p = 0.9968$) or a Mann–Whitney U test ($p = 0.9131$). Trend analyses likewise show no evidence of systematic change: a linear regression yields a slope indistinguishable from zero (p = 0.9886) and a Spearman test finds no monotonic movement ($\rho = 0.0929$, $p = 0.7420$). Results are similar for the Second-Price Sealed-Bid (SPSB) auction. The comparison between the first and last five rounds reveals no significant difference (means 0.1282 vs. 0.1447; t-test $p = 0.3335$; Mann–Whitney $p = 0.1948$), and neither a linear trend test (p = 0.3664) nor a Spearman correlation test ($\rho= 0.3964$, $p = 0.1435$) indicates meaningful temporal structure. Taken together, across all specifications, we find no statistical evidence of learning effects, behavioral drift, or systematic changes in bidding deviations over the 15 rounds. LLM agents' deviation magnitudes remain remarkably stable throughout both FPSB and SPSB environments.

\begin{figure}[H]
    \centering
    \includegraphics[scale=0.45]{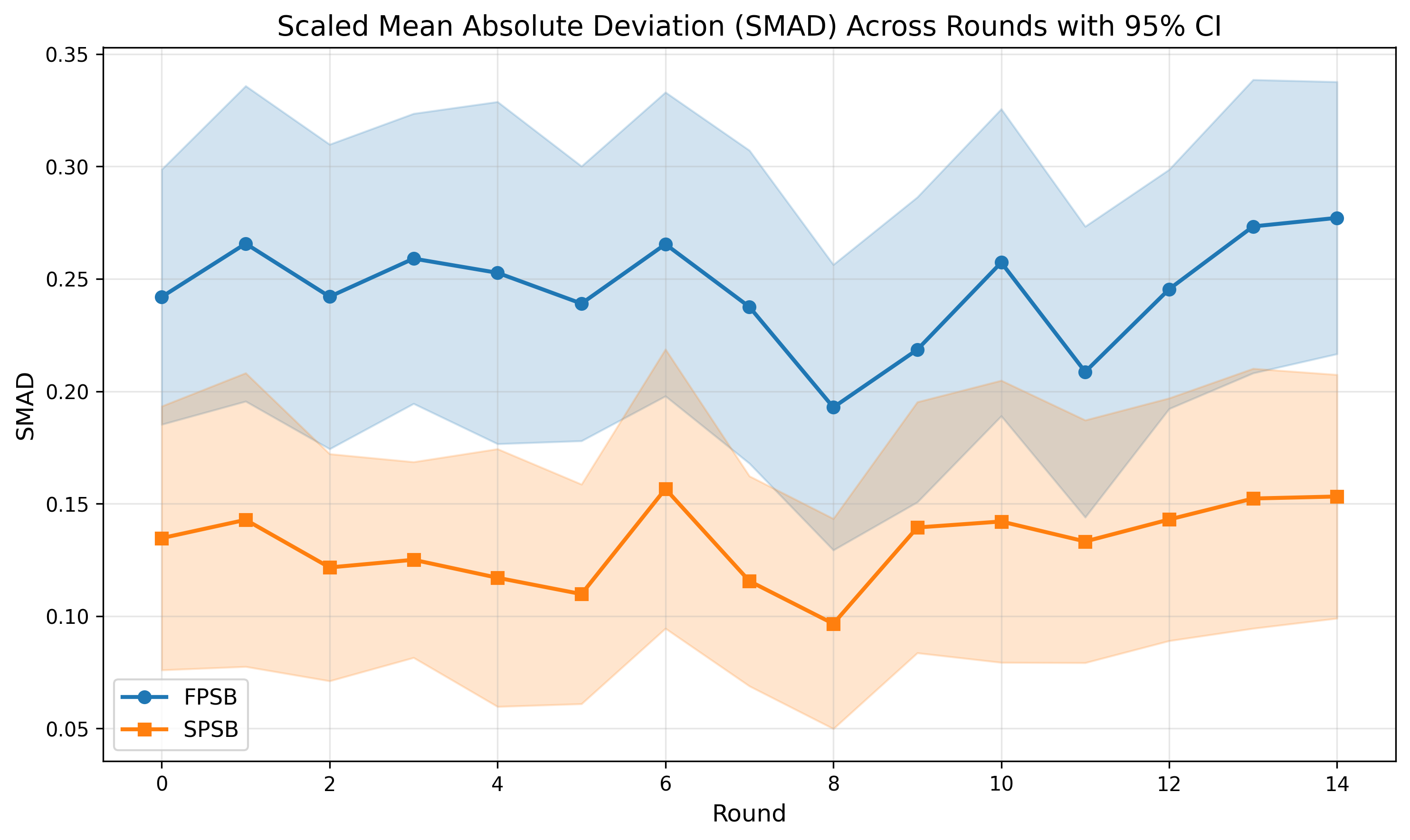}
    \caption{ Multiple-round simulation of First-Price and Second-Price sealed-bid auction with GPT-4o model (Temp=0.5). Y-axis stands for Scaled Mean Absolute Deviation (SMAD) between actual bidding and optimal bidding.
    \label{fig:multiple-round}}
\end{figure}

\section{Ablation on Models and Parameters}\label{app:ablation}
\subsection{Varying Temperature for GPT-4o Agent}\label{app:temperature}
We examined the effect of the temperature parameter (T $\in$ \{0.1, 0.5, 1.0\}) on GPT-4o's 
bidding behavior across seven auction formats. The temperature parameter controls the 
randomness of the model's outputs, with lower values producing more deterministic 
responses and higher values introducing more variability.

\rev{Temperature changes the sampling distribution of the same model without changing the mechanism, so it is a natural check on the ordinal claim. Table~\ref{tab:temperature_comparison} reports SMAD by format and temperature together with the ordering statistics of Section~\ref{sec:tiers}. The cross-mechanism ordering is robust to temperature. At every temperature, first-price is harder than second-price, the first-price common-value curse exceeds the second-price curse, the clock formats have the lowest deviations wherever they are measurable, and deviating bids in both second-price formats are predominantly underbids. Kendall's $\tau_b$ against the human ranking is positive in $100\%$ of joint bootstrap draws at every temperature, and the rank correlation of the SMAD vectors between temperatures is at least $+0.90$ (Spearman). The one exception is the order of the two closest formats, SP-APV and SPSB IPV, which matches the human order only at $T=0.5$. The clock cells at $T=0.1$ and $T=1.0$ have few usable bids and are marked in the table. The computation is in \texttt{analysis/temperature\_concordance.py}.}

\begin{figure}[H]
    \centering
\includegraphics[scale=0.35]{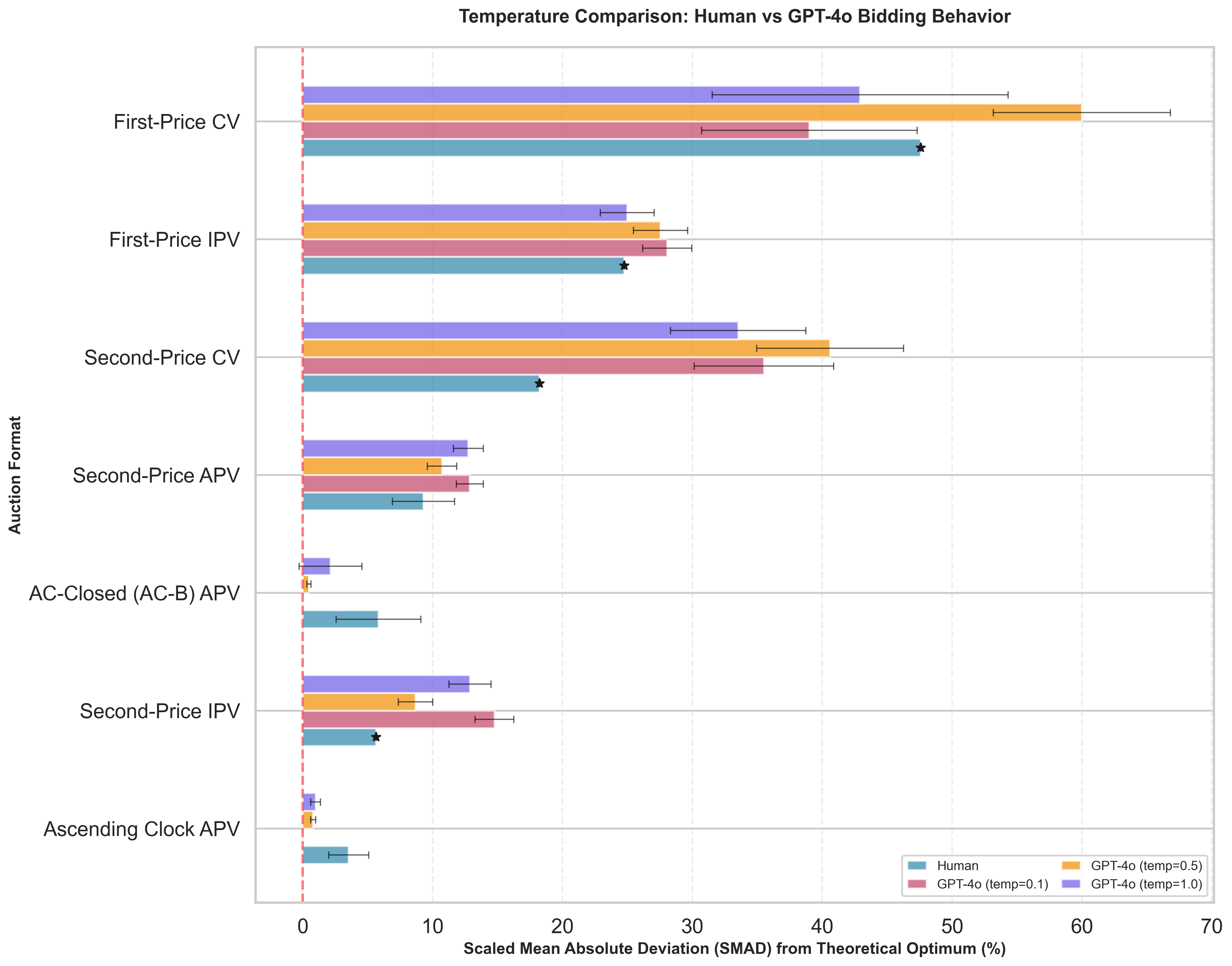}
    \caption{Comparison of the bidding behavior of human participants and GPT-4o across three
different temperature settings (T=0.1, 0.5, 1.0) in seven auction formats. The horizontal
bars represent the Scaled Mean Absolute Deviation (SMAD) from theoretical equilibrium,
with error bars indicating 95\% confidence intervals.
    \label{fig:temperature}}
\end{figure}

\begin{table}[htbp]
\centering
\footnotesize
\rev{
\begin{tabular}{@{}lccc@{}}
\toprule
 & $T=0.1$ & $T=0.5$ & $T=1.0$ \\
\midrule
FP-CV & 39.00 {\scriptsize (30)} & 59.99 {\scriptsize (96)} & 42.92 {\scriptsize (30)} \\
FPSB & 28.07 {\scriptsize (300)} & 27.55 {\scriptsize (285)} & 24.99 {\scriptsize (297)} \\
SP-CV & 35.51 {\scriptsize (100)} & 40.60 {\scriptsize (92)} & 33.53 {\scriptsize (99)} \\
SP-APV & 12.87 {\scriptsize (297)} & 10.73 {\scriptsize (291)} & 12.75 {\scriptsize (300)} \\
AC-B & 0.50$^{\dagger}$ {\scriptsize (4)} & 0.47 {\scriptsize (114)} & 2.15$^{\dagger}$ {\scriptsize (8)} \\
SPSB & 14.77 {\scriptsize (294)} & 8.69 {\scriptsize (270)} & 12.89 {\scriptsize (297)} \\
AC & n/a & 0.81 {\scriptsize (106)} & 1.00 {\scriptsize (52)} \\
\midrule
Kendall $\tau_b$ vs human (3 sealed) & $+0.33$ & $+1.00$ & $+0.33$ \\
 & {\scriptsize [$+0.33$, $+1.00$]} & {\scriptsize [$+1.00$, $+1.00$]} & {\scriptsize [$+0.33$, $+1.00$]} \\
\addlinespace
Kendall $\tau_b$ vs human (all $n\ge30$) & $+0.33$ & $+0.60$ & $+0.67$ \\
 & {\scriptsize [$+0.33$, $+1.00$], $k{=}3$} & {\scriptsize [$+0.40$, $+1.00$], $k{=}5$} & {\scriptsize [$+0.67$, $+1.00$], $k{=}4$} \\
\addlinespace
FPSB $>$ SPSB & yes {\scriptsize (100\%)} & yes {\scriptsize (100\%)} & yes {\scriptsize (100\%)} \\
FP-CV $>$ SP-CV & yes {\scriptsize (76\%)} & yes {\scriptsize (100\%)} & yes {\scriptsize (93\%)} \\
SPSB IPV: under-share of deviating bids & 87\% {\scriptsize (within 17\%)} & 81\% {\scriptsize (within 31\%)} & 88\% {\scriptsize (within 19\%)} \\
SP-APV: under-share of deviating bids & 90\% {\scriptsize (within 3\%)} & 92\% {\scriptsize (within 9\%)} & 85\% {\scriptsize (within 9\%)} \\
\bottomrule
\end{tabular}
}
\caption{\rev{Temperature effects on GPT-4o bidding. Top: SMAD (\%) by format with the number of usable bids (common value: auction rounds) in parentheses; $^{\dagger}$: fewer than 30 usable bids; n/a: no data. Bottom: the ordering and direction statistics at each temperature; Kendall's $\tau_b$ is against the reconstructed human ranking with joint 95\% bootstrap intervals.}}
\label{tab:temperature_comparison}
\end{table}

\subsection{Varying the Base Models for LLM bidding}\label{app:models}

\begin{figure}[H]
    \centering
\includegraphics[width=\linewidth]{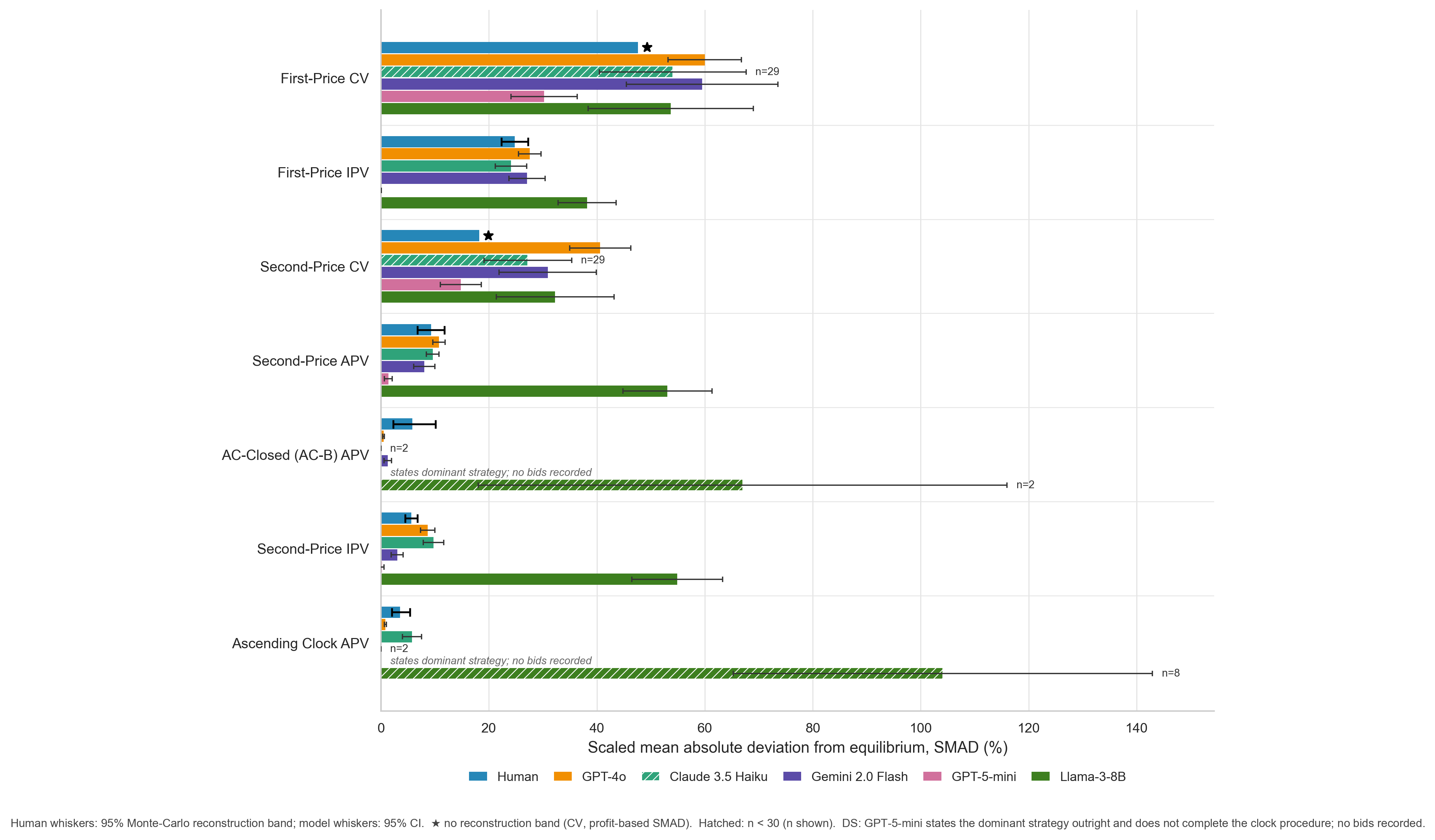}
    \caption{\rev{Bidding behavior of human participants (reconstructed; whiskers are 95\% reconstruction bands) and of the five LLMs (GPT-4o, Claude~3.5 Haiku, Gemini~2.0 Flash, GPT-5-mini, Llama-3-8B) in seven auction formats. Bars are SMAD from the theoretical benchmark with 95\% confidence intervals for LLM cells; cells with fewer than 30 usable bids are marked; in the clock formats GPT-5-mini states the dominant strategy and records no bids.}
    \label{fig:model_ablation}}
\end{figure}

\rev{Table~\ref{tab:model_comparison} in the main text reports SMAD for all five models; Figure~\ref{fig:model_ablation} shows the same cells graphically, with 95\% confidence intervals for the LLM cells and reconstruction bands for the human anchors, and Table~\ref{tab:direction_models} reports the direction of deviating bids by model. Two cautions apply to the cross-model comparison. First, in the two ascending-clock cells GPT-5-mini states the dominant strategy outright and does not complete the tick-by-tick clock procedure in any run, so no bids are recorded; the cells are marked DS$^{\ddagger}$ and enter no statistic. The clock cells are thin for the other non-reference models: only Claude~3.5 Haiku in the open clock and Gemini~2.0 Flash in the closed clock have more than a handful of usable (non-winning) bids; the remaining clock cells have $n \le 8$ and are reported for completeness only. (An earlier version of this table rendered these cells as $0.00$.) Second, we do not test LLM cells against the reconstructed human anchors: the anchors are reconstructions of published moments, not samples, so a test that treats them as known would be circular.}

\rev{GPT-5-mini, a reasoning model, sits essentially at the theoretical benchmark in the private-value sealed-bid formats (SMAD below $1.4\%$) and is therefore uninformative about the human difficulty ordering: with no error to remove, there is nothing to rank. Llama-3-8B sits at the other extreme, with SMAD above $38\%$ in every private-value format, and its ordering of formats is inverted relative to humans. Claude~3.5 Haiku and Gemini~2.0 Flash occupy the intermediate range in which the human ordering is reproduced (Appendix~\ref{app:concordance}). In the common-value formats all five models, like humans, deviate more under first-price than under second-price rules.}

\begin{table}[htbp]
\centering
\small
\rev{\resizebox{\linewidth}{!}{
\begin{tabular}{@{}l*{6}{c}@{}}
\toprule
Format & Human & GPT-4o & Claude 3.5 Haiku & Gemini 2.0 Flash & GPT-5-mini & Llama-3-8B \\
\midrule
FPSB IPV & 32 / 68 & 14 / 86 & 2 / 98 & 10 / 90 & n/a & 82 / 18 \\
 & {\scriptsize within 6\%} & {\scriptsize within 3\%, n=285} & {\scriptsize within 1\%, n=90} & {\scriptsize within 1\%, n=90} & {\scriptsize within 100\%, n=87 (no deviating bids)} & {\scriptsize within 1\%, n=90} \\
\addlinespace
SPSB IPV & 8 / 92 & 81 / 19 & 33 / 67 & 100 / 0 & 100 / 0 & 94 / 6 \\
 & {\scriptsize within 27\%} & {\scriptsize within 31\%, n=270} & {\scriptsize within 27\%, n=90} & {\scriptsize within 74\%, n=87} & {\scriptsize within 99\%, n=87} & {\scriptsize within 4\%, n=90} \\
\addlinespace
SP-APV & 20 / 80 & 92 / 8 & 70 / 30 & 100 / 0 & 100 / 0 & 93 / 7 \\
 & {\scriptsize within 50\%} & {\scriptsize within 9\%, n=291} & {\scriptsize within 3\%, n=90} & {\scriptsize within 30\%, n=87} & {\scriptsize within 83\%, n=87} & {\scriptsize within 9\%, n=90} \\
\addlinespace
AC & 45 / 55 & 50 / 50 & 21 / 79 & n/a$^{\dagger}$ & DS$^{\ddagger}$ & 88 / 12$^{\dagger}$ \\
 & {\scriptsize within 67\%} & {\scriptsize within 72\%, n=106} & {\scriptsize within 7\%, n=46} & {\scriptsize within 100\%, n=2 (no deviating bids)} &  & {\scriptsize within 0\%, n=8} \\
\addlinespace
AC-B & 29 / 71 & 37 / 63 & n/a$^{\dagger}$ & 88 / 12 & DS$^{\ddagger}$ & 100 / 0$^{\dagger}$ \\
 & {\scriptsize within 44\%} & {\scriptsize within 83\%, n=114} & {\scriptsize within 100\%, n=2 (no deviating bids)} & {\scriptsize within 63\%, n=46} &  & {\scriptsize within 0\%, n=2} \\
\bottomrule
\end{tabular}
}}
\caption{\rev{Direction of deviating bids by model and format: percentage of underbids / overbids among bids outside the $\pm 2\%$ band around the theoretical benchmark, with the within-band share and $n$ below. $^{\dagger}$: fewer than 30 usable bids; DS$^{\ddagger}$: GPT-5-mini states the dominant strategy outright and does not complete the clock procedure; no bids recorded.}}
\label{tab:direction_models}
\end{table}

\newpage

\section{\texorpdfstring{\rev{Rank Concordance Details}}{Rank Concordance Details}}\label{app:concordance}

\rev{Table~\ref{tab:concordance} reports Kendall's $\tau_b$ between the reconstructed-human SMAD ranking and each model's SMAD ranking. The five in-scope formats are FPSB IPV, SPSB IPV, SP-APV, AC, and AC-B, the private-value formats whose benchmark is on a common scale; the common-value formats use a profit-based deviation and third-price uses $N=5$, so neither enters the ranking. For each model we report $\tau_b$ over the three sealed-bid formats (always available) and over all formats with at least 30 usable bids. Uncertainty comes from three layers with $B=2000$ draws: (a) resampling LLM bids within each cell with the human ranking fixed; (b) re-drawing the human reconstruction within the admissible parameter ranges of Appendix~\ref{app:bands} with the LLM ranking fixed; (c) both jointly. For GPT-4o over five formats, $\tau_b = 0.60$ arises from 8 concordant and 2 discordant pairs, both involving AC-B, whose human anchor sits between SPSB IPV and SP-APV while GPT-4o places both clock formats at the bottom; the estimate is unchanged in $98.7\%$ of LLM-resampling draws and positive in $100\%$ of joint draws, with first-price ranked hardest on both sides in every draw. Claude~3.5 Haiku's single discordance over the sealed formats is the near-tie between its SPSB IPV ($9.72\%$) and SP-APV ($9.58\%$) cells, which flips under resampling in about $45\%$ of draws, so its interval reaches $1.00$. GPT-5-mini's negative point estimate reflects small differences between near-equilibrium bids, including two cells at $0.08\%$ and $0.20\%$. Llama-3-8B's ordering is inverted in all but one of $2000$ joint draws. With three formats $\tau_b$ takes values in $\{-1, -\tfrac13, \tfrac13, 1\}$, and the human-side layer is nearly degenerate over the sealed formats, so the three-format intervals are driven by LLM resampling. Two caveats: human-side uncertainty is reconstruction uncertainty, not sampling error; and LLM bids are treated as i.i.d.\ within cells. The computation is in \texttt{analysis/difficulty\_concordance\_models.py}.}


\section{\texorpdfstring{\rev{Winner's Curse: Details by Auction Format and Number of Bidders}}{Winner's Curse: Details by Auction Format and Number of Bidders}}\label{app:winners_curse}

\rev{This appendix collects the GPT-4o common-value results summarized in Section~\ref{sec:tier_orderings}, following the design of \citet{kagel1986winner}.}

Common-value auctions provide a particularly sharp test of whether bidders correctly condition on the adverse selection implicit in winning. In the canonical experimental design of \citet{kagel1986winner}, each bidder observes a noisy private signal $x$ about a common underlying value $x_0$. Winning is therefore ``bad news'': conditional on having the highest signal among $n$ bidders, the expected common value is strictly below $x$, with the gap widening in $n$.\footnote{In the standard symmetric-information design, $\mathbb{E}[x_0 \mid X=x_{1:n}] = x - \frac{n-1}{n+1}\varepsilon$ for signals away from boundary regions, where $\varepsilon$ controls signal noise. This highlights why competition exacerbates adverse selection: as $n$ rises, the winner must make a larger downward correction.} If bidders neglect (or incompletely correct for) this conditioning, they overbid relative to $\mathbb{E}[x_0\mid \text{win}]$ and the winner earns negative profits: the classic \emph{winner's curse}. Laboratory evidence documents both comparative statics emphasized in our setting: (i) winner losses become larger as the number of bidders increases, and (ii) institutions that reveal more information during the price discovery process attenuate the curse relative to sealed-bid first-price formats. For example, \citet{kagel1986winner} reports substantial negative profits conditional on winning that are markedly worse with more bidders (e.g., approximately $-\$1.68$ with $n=4$ versus $-\$3.68$ with $n=7$ in their benchmark conditions), and related work shows that the revenue advantage of ascending formats hinges on whether bidders have learned to avoid severe winner's-curse mistakes \citep{levin1996revenue}.

We evaluate whether LLM bidders exhibit analogous winner's-curse patterns in common-value \emph{first-price sealed-bid} (FPSB) auctions. Table~\ref{tab:winner_curse_llm} summarizes \emph{winner profits} in our simulations as we vary the number of bidders from 4 to 7 (holding the environment fixed). A clear winner’s-curse pattern emerges: winners lose money on average in every FPSB setting, and losses grow with competition (mean winner profit: $-2.54$ with 4 bidders, $-4.40$ with 5 bidders, $-5.46$ with 6 bidders, and $-5.55$ with 7 bidders). Consistent with this monotonic decline, the share of winners earning positive profits collapses from 24\% (4 bidders) to 0\% (7 bidders). One-sided tests strongly reject the hypothesis that winner profit is nonnegative in all FPSB treatments, confirming that LLM bidders fall prey to the winner’s curse\citep{kagel1986winner}.

To benchmark the role of auction format, we also simulate a common-value \emph{second-price sealed-bid} (SPSB) environment under the same information structure. The same comparative-static forces appear, but the curse is attenuated: with 4 bidders, winners earn positive profits on average (mean $=1.30$), profits are near zero with 5 bidders, and winner losses become statistically significant only once competition is sufficiently intense ($n\ge 6$). Thus, while both formats exhibit adverse-selection pressures that grow in $n$, LLM behavior displays a substantially stronger winner’s curse in FPSB than in SPSB, mirroring the qualitative ranking emphasized by the experimental literature\citep{levin1996revenue}.

\begin{table}[H]
\centering
\begin{tabular}{lccccc}
\toprule
\textbf{Auction} & \textbf{Bidders} & \textbf{Mean (Std.)} & \textbf{Median} & \textbf{\% Positive} & \textbf{$p$ (Profit $<0$)} \\
\midrule
FPSB & 4 & $-2.54$ (3.87) & $-3.00$ & 24.0\% & $<0.001^{***}$ \\
FPSB & 5 & $-4.40$ (3.63) & $-4.00$ & 6.0\%  & $<0.001^{***}$ \\
FPSB & 6 & $-5.46$ (3.82) & $-5.05$ & 4.0\%  & $<0.001^{***}$ \\
FPSB & 7 & $-5.55$ (2.93) & $-5.30$ & 0.0\%  & $<0.001^{***}$ \\
\midrule
SPSB & 4 & $1.30$ (4.87)  & $1.00$  & 58.0\% & 0.967 \\
SPSB & 5 & $0.07$ (4.35)  & $0.00$  & 46.0\% & 0.542 \\
SPSB & 6 & $-2.82$ (3.50) & $-1.75$ & 18.0\% & $<0.001^{***}$ \\
SPSB & 7 & $-2.83$ (2.78) & $-2.00$ & 12.0\% & $<0.001^{***}$ \\
\bottomrule
\end{tabular}
\caption{Winner profit in common-value auctions with LLM bidders. The last column reports one-sided $p$-values for the hypothesis that winner profit is negative. *** denotes $p < 0.001$.}
\label{tab:winner_curse_llm}
\end{table}

\section{\texorpdfstring{\rev{Robustness of the Clock-Framing Effect: An Additional Model Family}}{Robustness of the Clock-Framing Effect: An Additional Model Family}}\label{app:clock_robustness}

\begin{figure}[H]
    \centering
    \includegraphics[width=0.45\linewidth]{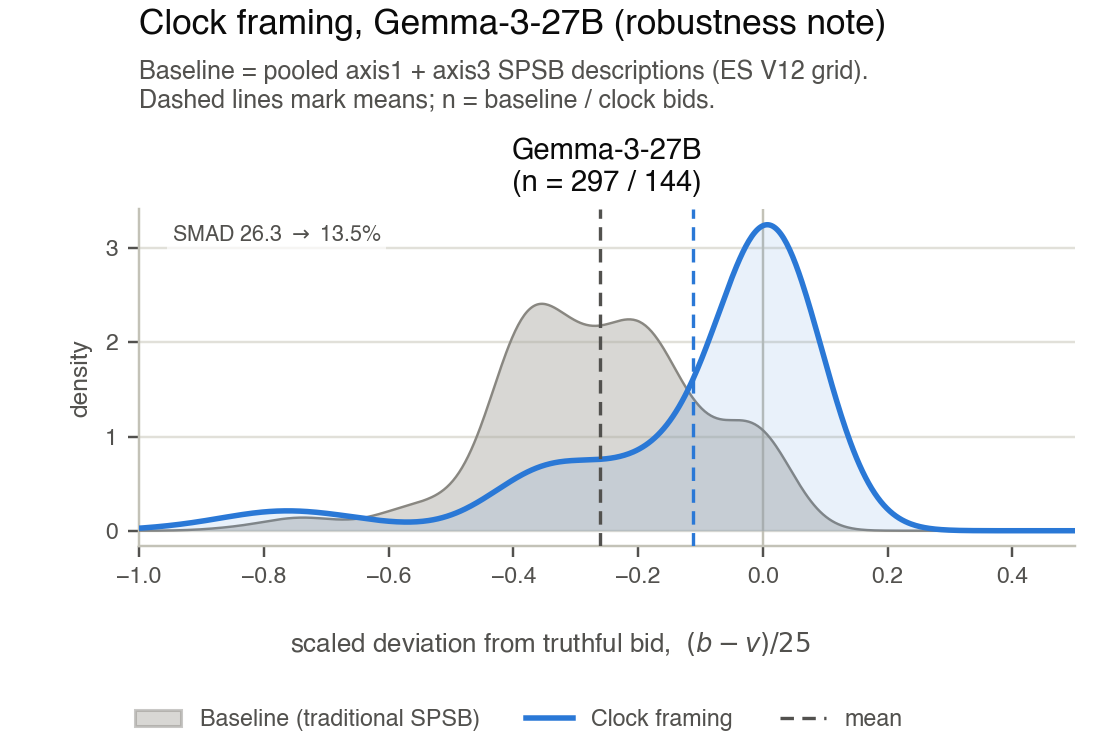}
    \caption{\rev{Clock framing of the second-price sealed-bid auction for Gemma-3-27B: distribution of per-bid scaled deviations $(b-v)/\mathbb{E}[b^\star]$ under the baseline description (grey) and the clock framing (blue), with means marked.}}
    \label{fig:clock_framing_gemma}
\end{figure}

\rev{Section~\ref{session:Intervention} tests the clock framing on the three non-reasoning large model families of the main panel. As a robustness check we repeat the test on Gemma-3-27B, an open-weights model outside the main panel whose baseline cells were run as part of the same intervention battery with identical rule templates (Appendix~\ref{app:intervention}); the baseline cell pools two independently run instances of the standard second-price description. Gemma's baseline second-price play is the least truthful of the four families ($26.3\%$ SMAD; $6\%$ of bids within $\pm2\%$ of value) but well above the noise floor of Llama-3-8B, so the ordinal argument predicts a gain. The clock framing lowers Gemma's SMAD from $26.3\%$ to $13.5\%$ (Mann--Whitney $p<0.001$) and raises the share of bids within $\pm2\%$ of value from $6\%$ to $22\%$ (Table~\ref{tab:clock_framing_gemma}, Figure~\ref{fig:clock_framing_gemma}). Across the four families, the clock framing therefore helps in three and fails only for the family closest to truthful play at baseline. Two wording differences between the earlier GPT-4o run used in a previous version of this paper and the cells reported here should be noted: the earlier run stated that bids must lie between \$0 and the bidder's value in \$0.1 increments (the current templates say only that bids must be in \$0.1 increments), and it limited the bidder's written plan to 50 rather than 100 words. The computation is in \texttt{scripts/plots/interventions\_models.py}.}

\begin{table}[htbp]
\centering
\small
\rev{\resizebox{\linewidth}{!}{\begin{tabular}{l rr rr c l c}
\toprule
 & \multicolumn{2}{c}{Baseline} & \multicolumn{2}{c}{Clock framing} & within $\pm2\%$ & & matches \\
\cmidrule(lr){2-3}\cmidrule(lr){4-5}
Model & SMAD & mean dev & SMAD & mean dev & baseline $\to$ clock & $\Delta$ (MW) & human sign? \\
\midrule
Gemma-3-27B & 26.3\% & $-0.261$ & 13.5\% & $-0.112$ & 6\% $\to$ 22\% & $\downarrow$ {\scriptsize $p<0.001$} & $\surd$ \\
\bottomrule
\end{tabular}
}}
\caption{\rev{\textbf{Clock framing of the second-price sealed-bid auction for Gemma-3-27B.} Columns as in Table~\ref{tab:clock_framing_main}.}}
\label{tab:clock_framing_gemma}
\end{table}

\section{\texorpdfstring{\rev{Additional Results for the eBay-Style Environment}}{Additional Results for the eBay-Style Environment}}\label{app:ebay_extra}

\begin{figure}[H]
    \centering
    \includegraphics[width=0.7\linewidth]{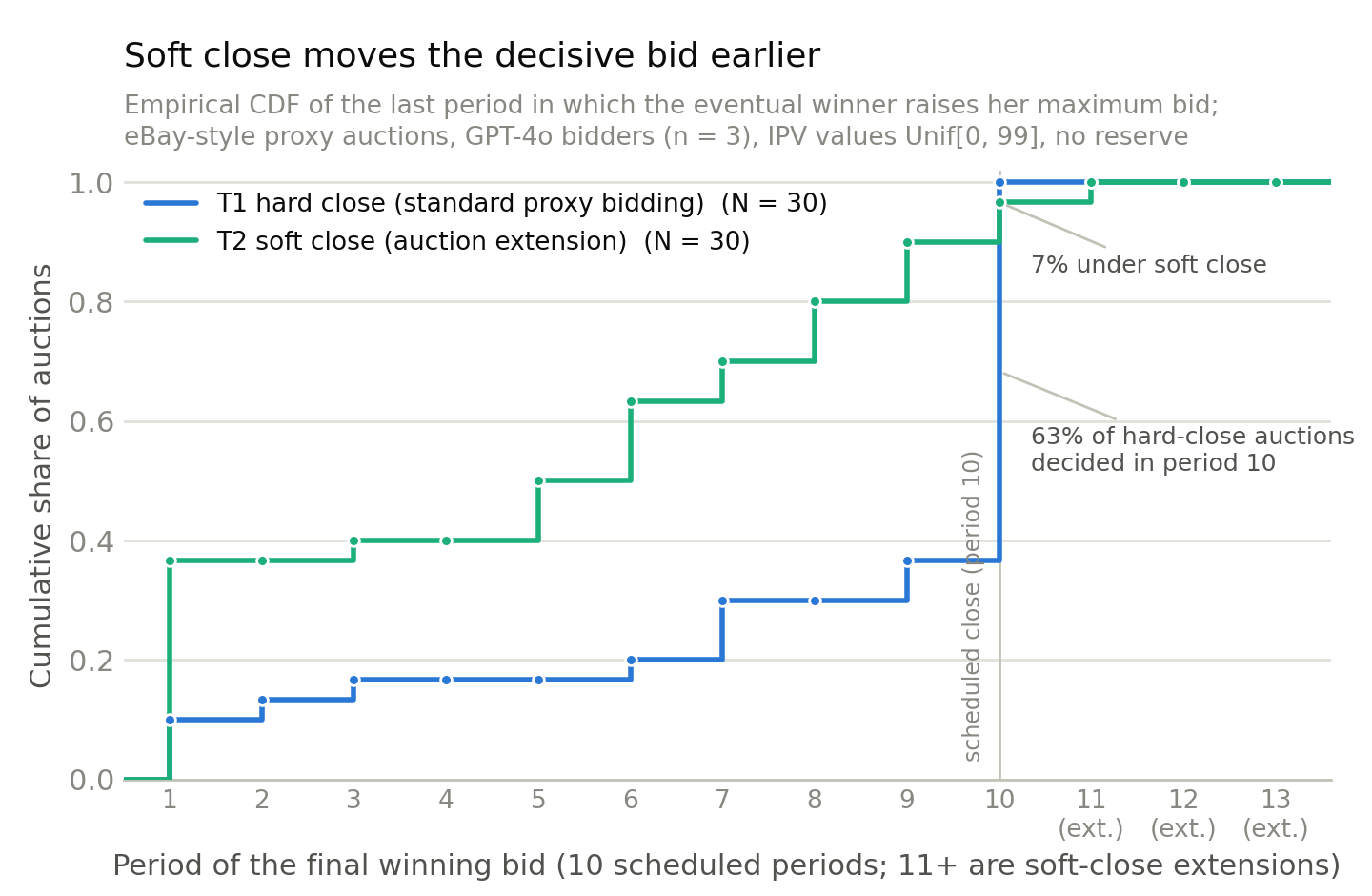}
    \caption{\rev{\textbf{CDF of final winning-bid timing under the two closing rules, eBay-style auctions, IPV environment.} The cumulative distribution function of the period of the final winning bid under the standard hard close (T1) and the soft-close extension rule (T2). The soft close shifts the final winning bid substantially earlier, reproducing the ordering of last-minute bidding across closing rules documented for eBay and early Amazon auctions \citep{roth2002last, ockenfels2006late}.}
    \label{fig:ebay_cdf_final_winning_bid}}
\end{figure}

\rev{Figure~\ref{fig:ebay_cdf_final_winning_bid} reports the cumulative distribution of the final winning-bid period under the two closing rules of Section~\ref{sec:ebay_market}, complementing the histograms in Figure~\ref{fig:final_winning_bid_shift}.}

\section{Auction Rule Prompts}\label{app:prompt}
Below, we report the prompts used for each section.

\subsection{Independent Private Values Prompts}\label{sec:ipv-prompts}
Following the simulation procedure outlined in Section~\ref{app:process}, we employ the following prompts for the four auctions tested in this section. For each prompt, only the third paragraph (describing the allocation rule and payment) differ. For the FPSB, we describe the auction format as follows:
\begin{figure}
\centering
\begin{minipage}{0.95\textwidth}
\begin{lstlisting}
In this game, you will participate in an auction for a prize against {{num_bidders}} other bidders. You will play this game for {{n}} rounds.
At the start of each round, bidders will see their value for the prize, randomly drawn between $0 and ${{private}}, with all values equally likely.
After learning your value, you will submit a bid privately at the same time as the other bidders. Bids must be in ${{increment}} increments.
The highest bidder wins the prize and pays their bid amount. If you win, your earnings will increase by your value for the prize, and decrease by your bid. If you don't win, your earnings will remain unchanged.
After each auction, we will display all bids. Ties for the highest bid will be resolved randomly.
\end{lstlisting}
\end{minipage}
\label{fig:sealbid_prompt-1}
\end{figure}

For the SPSB, we describe the auction format as follows:
\begin{figure}
\centering
\begin{minipage}{0.95\textwidth}
\begin{lstlisting}
In this game, you will participate in an auction for a prize against {{num_bidders}} other bidders. You will play this game for {{n}} rounds.
At the start of each round, bidders will see their value for the prize, randomly drawn between $0 and ${{private}}, with all values equally likely.
After learning your value, you will submit a bid privately at the same time as the other bidders. Bids must be in ${{increment}} increments.
The highest bidder wins the prize and pays the second-highest bid. If you win, your earnings will increase by your value for the prize, and decrease by the second-highest bid. If you don't win, your earnings will remain unchanged.
After each auction, we will display all bids. Ties for the highest bid will be resolved randomly.
\end{lstlisting}
\end{minipage}
\label{fig:sealbid_prompt-2}
\end{figure}

For the TPSB, we describe the auction format as follows:
\begin{figure}
\centering
\begin{minipage}{0.95\textwidth}
\begin{lstlisting}
In this game, you will participate in an auction for a prize against {{num_bidders}} other bidders. You will play this game for {{n}} rounds.
At the start of each round, bidders will see their value for the prize, randomly drawn between $0 and ${{private}}, with all values equally likely.
After learning your value, you will submit a bid privately at the same time as the other bidders. Bids must be in ${{increment}} increments.
The highest bidder wins the prize and pays the third-highest bid. If you win, your earnings will increase by your value for the prize, and decrease by the third-highest bid. If you don't win, your earnings will remain unchanged. If there are less than three bids, no one will win the auction.
After each auction, we will display all bids. Ties for the highest bid will be resolved randomly.
\end{lstlisting}
\end{minipage}
\label{fig:sealbid_prompt-3}
\end{figure}

\subsection{Obviously Strategy-Proof Prompts}\label{sec:osp-prompts}
The sealed-bid prompt follows the style of the last section's prompts closely. The only difference is in the first two paragraphs, where agents are informed of the APV environment. For the SPSB, we describe the auction format as follows:
\begin{figure}
\centering
\begin{minipage}{0.95\textwidth}
\begin{lstlisting}
In this game, you will participate in an auction for a prize against {{num_bidders}} other bidders. You will play this game for {{n}} rounds.
At the start of each round, bidders will see their value for the prize. Your value for the prize will be calculated as follows:
1. First we will randomly draw a common value between {{common_low}} and {{common_high}}, with all values equally likely.
2. Then, for each bidder, a private taste adjustment will be drawn between 0 and {{private}}, with all values equally likely.
Your value for the prize is equal to the common value plus your private taste adjustment. You will not learn the common value or your private taste adjustment separately. This means that each person in your group may have a different value for the prize. However, if you have a high value, it is more likely that other people in your group have a high value.
After learning your value, you will submit a bid privately at the same time as the other bidders. Bids must be in ${{increment}} increments. 
The highest bidder wins the prize and pays the second-highest bid. If you win, your earnings will increase by your value for the prize, and decrease by the second-highest bid. If you don't win, your earnings will remain unchanged.
After each auction, we will display all bids. Ties for the highest bid will be resolved randomly.
\end{lstlisting}
\end{minipage}
\label{fig:sealbid_prompt-4}
\end{figure}

The implementation of the clock auction simulates a global clock price, with prices beginning at $0$ and increasing in $\$1$ increments. The clock auction format descriptions closely mirror the SPSB format above, but with the added description of the clock mechanism in the fourth and fifth paragraphs. 
In the AC auction, we describe the auction format as follows: 
\begin{figure}
\centering
\begin{minipage}{0.95\textwidth}
\begin{lstlisting}
In this game, you will participate in an auction for a prize against {{num_bidders}} other bidders. You will play this game for {{n}} rounds. 
At the start of each round, bidders will see their value for the prize. Your value for the prize will be calculated as follows:
1. First we will randomly draw a common value between {{common_low}} and {{common_high}}, with all values equally likely.
2. Then, for each bidder, a private taste adjustment will be drawn between 0 and {{private}}, with all values equally likely.
Your value for the prize is equal to the common value plus your private taste adjustment. You will not learn the common value or your private taste adjustment separately. This means that each person in your group may have a different value for the prize. However, if you have a high value, it is more likely that other people in your group have a high value.
The auction proceeds as follows: First, you will learn your value for the prize. Then, the auction will start. We will display a price to everyone in your group that starts at 0 and counts upwards in {{increment}} USD increments, up to a maximum of {{common_high + private}}. At any point, you can choose to leave the auction, and anytime a bidder leaves, we will broadcast that information to all the remaining bidders. 
When there is only one bidder left in the auction, that bidder will win the prize at the current price. If you win, your earnings will increase by your value for the prize, and decrease by the current price. If you don't win, your earnings will remain unchanged.
After each auction, we will display all bids. Ties for the highest bid will be resolved randomly.
\end{lstlisting}
\end{minipage}
\label{fig:sealbid_prompt-5}
\end{figure}

The AC-B auction description closely resembles the AC format, but with the explicit caveat that we will not notify bidders when competitors drop out.

\subsection{Common Value Auctions}

For the Second-Price Sealed-Bid auction in the Common Value setting, the rule is

\begin{figure}[H]
\centering
\begin{minipage}{0.95\textwidth}
\begin{lstlisting}
In this game, you will participate in an auction for a prize against {{num_bidders}} other bidders. You will play this game for {{n}} rounds. 
At the start of each round, bidders will see their perceived value for the prize- a noisy measurement of the true value of the prize. Your perceived value for the prize will be calculated as follows:
1. For each round, a common value will be drawn between {{common_low}} and {{common_high}}, with all values equally likely to be drawn.
2. For each person, a private noisy adjustment will be drawn between -{{private}} and {{private}}, with all values equally likely to be drawn.
We will tell you your perceived value, the sum of the common value and the private noise adjustment. However, everyone's true value for the prize is equal to the shared common value. 
After learning your perceived value, you will submit a bid privately at the same time as the other bidders. Bids must be in ${{increment}} increments. 
The highest bidder wins the prize and pays the second-highest bid. If you win, your earnings will increase by the true value for the prize, and decrease by the second-highest bid. If you don't win, your earnings will remain unchanged.
After each auction, we will display all bids. Ties for the highest bid will be resolved randomly.
\end{lstlisting}
\end{minipage}
\label{fig:sealbid_prompt-6}
\end{figure}

\subsection{Ebay-proxy prompt}
For the Standard eBay auction with proxy bidding (T1), the rule is:
\begin{figure}[H]
\centering
\begin{minipage}{0.95\textwidth}
\begin{lstlisting}
In this game, you will participate in an eBay auction for an item against {{num_bidders}} other bidders. This auction will last for {{num_rounds}} days. All dollar amounts in this game are in US Dollars ($).

Item Details:
Item Description: {{item_description}}
Item Condition: {{item_condition}}
Your Private Value: At the start of each round, bidders will see their value for the item, randomly drawn between $0 and ${{private}}, with all values equally likely. After learning your value for the item, you will submit a maximum bid. Bids must be in ${{increment}} increments.

Auction Format:
This is an eBay auction. The auction starts at ${{start_price}} and will last for {{num_rounds}} days. eBay uses "proxy bidding." This means that if you wish to enter the auction, you should submit your maximum bid, and eBay will automatically bid on your behalf, up to your maximum, only as much as necessary to maintain your position as the highest bidder. Each day you will see the current price and have the opportunity to increase your maximum bid. If you do not want to increase your maximum bid, then output HOLD.

You may place bids at any point during the auction, even on the final day. No bidder will know if they (or anyone else) is the last bidder on the final day.

The highest bidder wins the prize and pays the auction price at the time the auction's clock expired. If you win, your earnings will increase by your value for the prize, and decrease by the final auction price. If you don't win, your earnings will remain unchanged. Ties for the highest bid will be resolved randomly.
\end{lstlisting}
\end{minipage}
\label{fig:sealbid_prompt-7}
\end{figure}

For the eBay auction with a modified closing rule (T2), the updated closing rule in the prompt is

\begin{figure}[H]
\centering
\begin{minipage}{0.95\textwidth}
\begin{lstlisting}
This auction also has a closing rule. This means that if a new maximum bid is placed on the last day, the auction will automatically extend by another day. This extension will continue to be applied as long as new maximum bids are placed on the final day. No bidder will know if they (or anyone else) is the final bidder on the last day.
\end{lstlisting}
\end{minipage}
\label{fig:sealbid_prompt-8}
\end{figure}

\subsection{Intervention Prompt}\label{app:intervention}

%
%
We designed the clock framing intervention prompt based on \citet{breitmoser2022obviousness}.
\begin{figure}[H]
\centering
\begin{minipage}{0.95\textwidth}
\begin{lstlisting}
FIRST STAGE: Sealed Bid
    You will submit a sealed bid privately at the same time as the other bidders. This bid will serve as your automatic exit price in the next stage.  
SECOND STAGE: Ascending Clock (Simulation) 
    After the sealed bid stage, we will simulate an ascending clock auction.
    The clock will start at 0 and increase in increments of {increment}.
    The clock will display the current price. You will also see that there are a total of \{num$\_$bidders\} bidders participating, although you do not know other bidder's values.
    If the current price on the clock reaches or exceeds your sealed bid from the first stage, you will automatically exit the auction. The auction ends when only one bidder is left remaining in the second stage based on their bid from the first stage. 
    END OF AUCTION
\end{lstlisting}
\end{minipage}
\label{fig:sealbid_prompt-12}
\end{figure}

\end{document}